\documentclass[12pt,epsfig]{article}
\usepackage{amssymb,epsfig}

\makeatletter
\@addtoreset{equation}{section}

\newcommand{\bA}{{\bf A}}
\newcommand{\bB}{{\bf B}}
\newcommand{\bC}{{\bf C}}
\newcommand{\bF}{{\bf F}}
\newcommand{\bS}{{\bf S}}
\newcommand{\bZ}{{\bf Z}}
\newcommand{\bPhi}{{\bf \Phi}}
\newcommand{\ap}{{\alpha'}}

\newcommand{\cF}{{\cal F}}

\newcommand{\cR}{{\cal R}}

\newcommand{\vphi}{\varphi}

\newcommand{\nn}{\nonumber \\}

\newcommand{\tr}{{\rm tr}}

\newcommand{\Omeh}{{\hat \Omega}}

\newcommand{\xt}{{\widetilde x}}

\newcommand{\psit}{{\widetilde \psi}}

\def\be{\begin{equation}}
\def\ee{\end{equation}}
\def\beq{\begin{equation}}
\def\eeq{\end{equation}}
\def\bea{\begin{eqnarray}}
\def\eea{\end{eqnarray}}
\begin{document}

\begin{titlepage}
\hfill
\vbox{
    \halign{#\hfil         \cr
%           CERN-TH/2002-345 \cr
           TAUP-7265-04\cr
           hep-th/0403256  \cr
           } %  end of \halign
      }  % end of \vbox
\vspace*{20mm}

\begin{center}
{\Large {\bf Topological B-Model, Matrix Models, $\hat{c}=1$
    Strings\\ and Quiver Gauge Theories}\\} 
\vspace*{15mm}

{\sc Harald Ita},
\footnote{e-mail: {\tt ita@post.tau.ac.il}} 
{\sc Harald Nieder},
\footnote{e-mail: {\tt harald@post.tau.ac.il}} 
{\sc Yaron Oz} 
\footnote{e-mail: {\tt yaronoz@post.tau.ac.il}}
and 
{\sc Tadakatsu Sakai}
\footnote{e-mail: {\tt tsakai@post.tau.ac.il}}\\
\vspace*{1cm} 
{\it Raymond and Beverly Sackler Faculty of Exact Sciences\\
School of Physics and Astronomy\\
Tel-Aviv University , Ramat-Aviv 69978, Israel\\}

\end{center}

\vspace*{8mm}

\begin{abstract}

We study topological and integrable aspects
of $\hat{c}=1$ strings. 
We consider 
the circle line theories 0A and 0B at particular radii, and 
the super affine theories at their self-dual radii.
We construct their ground rings, identify them with certain 
quotients of the conifold, and
suggest  topological B-model descriptions.
We consider the partition functions, correlators and Ward 
identities, and construct a Kontsevich-like matrix model.
We then study all these aspects via the topological B-model
description.  
Finally, we analyse the corresponding Dijkgraaf-Vafa type matrix models
and quiver gauge theories.

\end{abstract}
\vskip 1cm

March  2004

\end{titlepage}

\setcounter{footnote}{0}

\section{Introduction}

In this paper we will consider the connections between four types of systems: non-critical strings, 
topological field theories, matrix models and ${\cal N}=1$ supersymmetric gauge theories in four dimensions
(for a recent study see \cite{Aganagic:2003qj}).
The most studied example in this context is 
the non-critical $c=1$ bosonic string at the self-dual radius. It has been argued to be equivalent
to the topological B-model on the deformed conifold, and 
the corresponding matrix model and gauge theory are described by the $\hat{A}_1$ quiver diagram
\cite{Dijkgraaf:2003xk}.

A basic ingredient in establishing the connections between these different systems is a 
commutative and associative
ring structure. For the $c=1$ bosonic string it is the ring of BRST invariant operators
with zero dimension and zero ghost number. Its defining relation for vanishing cosmological
constant $\mu$  is the conifold equation \cite{Witten:1991zd}.
It has been argued that when $\mu \neq 0$, the defining equation
describes the deformed conifold.
This ground ring relation has a corresponding chiral ring relation in the gauge theory
\cite{Dijkgraaf:2003xk}.

The partition function of the $c=1$ bosonic string as a function of the cosmological constant
$\mu$ is matched with that of the topological B-model on the deformed conifold, with $\mu$ being the deformation
parameter, and with the partition function of the $\hat{A}_{1}$ quiver matrix model,
 where $\mu$ is identified with $g_s N$, $N$ being the size of the matrix.
In the four-dimensional 
gauge theory $\mu$ corresponds to the glueball superfield $S$, and the partition function
is identified with the glueball F-terms \cite{Dijkgraaf:2003xk}.

Since the  $c=1$ bosonic string is well defined only perturbatively, the above connections are 
perturbative.
It is of interest to ask whether there 
are other examples of such connections, where the non-critical string is
well defined non-perturbatively.
Natural candidates are the fermionic
$\hat{c}=1$ strings, which have been introduced recently in \cite{Takayanagi:2003sm, Douglas:2003up}.

In this paper we will consider two families: 
the circle line theories 0A and 0B at particular radii, and 
the super affine theories at their self-dual radii.
Our aim is to study aspects of the connections between these non-critical strings and 
topological field theories, matrix models,
 and ${\cal N}=1$ supersymmetric gauge theories in four dimensions.
We will first
construct their ground rings, and identify them with certain $Z_2$ quotients of the conifold.
We will then consider integrable and topological aspects
of the theories. We will look at the partition functions, correlators and Ward 
identities, and construct a $0$A Kontsevich-like matrix model description.
Finally, we will
analyse the corresponding Dijkgraaf-Vafa type matrix models
and quiver gauge theories.

The paper is organized as follows.
In section 2 we will review some basics of $\hat{c} = 1$ strings, and identify candidate radii for a possible topological
description.
In section 3 we will construct the ground rings and propose topological B-model descriptions. 
In section 4 we will consider the partition functions and correlators.
In section 5 we will discuss the integrable structure and the Kontsevich-like matrix model.
In section 6 we will consider the topological B-model description.
In section 7 we will analyse the corresponding quiver gauge theories and matrix models. 
Section 8 is devoted to a discussion.

\section{$\hat{c} = 1$ Strings}

Fermionic strings are described by $N=1$ supersymmetric worldsheet 
field theories coupled to worldsheet supergravity.
Consider a 
nonchiral GSO projection which gives type 0
string theory.
There are two distinct choices, depending on how the
worldsheet fermion number $(-1)^F$ symmetry is realized in the closed string 
R-R sector, they are called $0$A and $0$B.
In both theories there 
are no $(NS,R)$ or $(R,NS)$ sectors, and therefore no spacetime fermions.
We will be interested in two-dimensional string theories.
Consider the spectrum of these theories.
In the NS-NS sector one has the (massless) tachyon physical field,
with additional discrete states at special values of the
momenta.
In the  R-R sector there are two vector fields $C_1$,
$\tilde C_1$ in type 0A, a scalar
$C_0$ 
in type 0B theory, and additional discrete states.

Let us compactify the theory on a circle of radius $R$.
We can choose as a matter system a free scalar superfield 
\beq
X = x+i\theta\chi+i\bar\theta\bar\chi+i\theta \bar\theta G \ ,
\eeq
or
any other $\hat{c}=1$ superconformal field theory.
We will consider 
two lines of theories parametrized by the radius $R$
of compactification of the lowest component of $X$:
The `circle' theories 0A and 0B, and 
the `super affine' theories, obtained from the
circle by modding out by a $Z_2$ symmetry $(-)^{\bF_L} e^{i\pi p}$
where $(-)^{\bF_L}=1\;(-1)$ on NS-NS (R-R) momentum states, and $p$ is the momentum $k=\frac{p}{R}$ \cite{Dixon}.

The vertex operators take the form \cite{Douglas:2003up}:
\begin{itemize}

\item{} NS-NS sector (in the $(-1,-1)$ picture):
\beq
  T_{k}^{(\pm)}=c\bar c\; \exp\left[-(\varphi+\bar\varphi)
    +ik(x_L+x_R)+(1\mp k)\phi\right] \ .
\eeq

\item{} RR sector 
(in the $(-\frac{1}{2},-\frac{1}{2})$ picture): 
\beq
  V_{k}^{(\pm)}=c\bar c\;\exp[
    -{\varphi+\bar\varphi\over2}\mp{i\over2}(H+\bar H) 
    +ik(x_L+x_R)+(1\mp k)\phi ] \ .
\eeq

\end{itemize}

Winding modes $\hat T_k$,
$\hat V_k$ take  the same form with 
$x_R, \bar H\rightarrow-x_R, -\bar H$, $k={wR\over2}$.
$\phi$ is the lowest component of the super Liouville field $\Phi$
\beq
\Phi = \phi + i\theta\psi + i \bar \theta
\bar \psi + i\theta\bar \theta F \ ,
\eeq
and 
$\psi+i\chi = {\sqrt2}\,e^{iH}$.

Imposing the GSO projection one gets \cite{Douglas:2003up}:

\begin{itemize}

\item{} type 0A:
In the NS-NS sector we have the tachyon field
momentum states with $k={n\over R}$, and winding states
with $k={wR\over2}$, 
where here and in the following $n,w$ take integer values. In the R-R sector we
remain with 
the winding modes with $k={wR\over2}$ and  project out
the momentum modes. 
%Here we set $\alpha^{\prime}=2$.

\item{} type 0B:
In the NS-NS sector we have the tachyon field
momentum states with $k={n\over R}$, and winding states
with $k={wR\over2}$. In the R-R sector we remain with 
the momentum modes with $k={n\over R}$, while projecting out
the winding modes.

\item{} Super affine 0A:  In the NS-NS
sector we have momentum modes tachyon  with $k={2n\over R}$
and tachyon winding modes with  
$k={wR\over2}$. The R-R states are projected  out.

\item{} Superaffine  0B: In the NS-NS
sector we have tachyon momentum modes  with $k={2n\over R}$
and tachyon winding modes with $k={wR\over2}$.
In the R-R sector we have 
momentum modes 
with $k={2n+1\over R}$
and windings 
with $k=(w+\frac{1}{2}){R\over2}$.

\end{itemize}

Type 0A and Type 0B are T-dual
to each other under
$R\rightarrow{\alpha'\over R}$.
The super affine theories are
self dual under
$R\rightarrow{2\alpha'\over R}$.

The torus partition functions of these theories read \cite{Douglas:2003up}:
\bea
F_{\rm 0B} =  - {\ln\mu\over 12 \sqrt 2}
\left ( {R\over \sqrt{\ap}} + 2 {\sqrt{\ap}\over R} \right ) \ , \nonumber\\
F_{\rm 0A} = - {\ln\mu\over 12 \sqrt 2}
\left ( 2 {R\over \sqrt{\ap}} + {\sqrt{\ap}\over R} \right ) \ ,
\eea
where $\mu$ corresponds to the (renormalized) cosmological constant in the super Liouville
action.
T-duality interchanges the two. For the superaffine theories we have
\bea
F_{\rm 0A}^{\rm super-affine} =- {
\ln\mu\over 12  } \left({ R\over \sqrt{2\ap}}+{ \sqrt{2 \ap}\over
R}\right) \ , \nonumber\\
F_{\rm 0B}^{\rm super-affine} =- {
\ln\mu\over 24}  \left({ R\over \sqrt{2\ap}}+{ \sqrt{2 \ap}\over R}\right) \ .
\eea
These suggest that we should consider as candidate ``topological points''
the radii\footnote{An interesting feature of the
radii $R_{\rm 0A}$ and $R_{\rm 0B}$ is that for these values the
non-perturbative contributions due to
D-instantons and wrapped D$0$-branes coincide. For a recent discussion
of these effects in the CFT and the matrix model approach 
see \cite{Martinec:2003ka}\cite{Alexandrov:2003nn}\cite{Alexandrov:2003un}.}
\beq
R_{\rm 0A}= \frac{l_s}{\sqrt{2}},~~~~~R_{\rm 0B}= l_s\sqrt{2},~~~~
R_{\rm super-affine}=l_s\sqrt{2} \ .
\label{topp}
\eeq
We will argue that at these points the non-critical $\hat{c}=1$ string theories have
a description as a topological B-model on a Calabi-Yau 3-fold, in a sense that
will be made more precise later.

These radii are chosen in analogy with  the $c=1$ bosonic string at the self-dual radius
$R^{\rm c=1}_{\rm self-dual} = l_s$, 
which has a topological
description.
At these radii the winding modes contribute exactly as the momentum 
modes to the torus partition function.
Note, however, that while the  $c=1$ bosonic string and the superaffine theories are 
self-dual at $R_{\rm super-affine}=l_s\sqrt{2}$, type $0$A and $0$B are not, but are
rather T-dual to each other. The results for the torus partition function
at the special radii (\ref{topp}) are displayed in table \ref{table1}.

%At the special radii (\ref{topp}) we have

\begin{table}[h] 
\hspace{2cm}
\begin{tabular}{||l|l||l|l||}\hline $F_{\rm c=1}(\rm R_{self-dual})
$&$-\frac{1}{12}\ln\mu$&&\\\hline
$F_{\rm 0A}(\sqrt{\alpha'/2})$&$-\frac{1}{6}\ln\mu$&$F_{\rm 0B}
(2\sqrt{\alpha'/2})$&$-\frac{1}{6}\ln\mu$\\\hline
$F_{\rm 0A}^{\rm super-aff.}(R_{\rm self-dual})$&$-\frac{1}{6}\ln\mu$&&\\\hline
$F_{\rm 0B}^{\rm super-aff.}(R_{\rm self-dual})$&$-\frac{1}{12}\ln\mu$&&\\\hline
\end{tabular}\\\\
\caption{Torus partition function at the special radii (\ref{topp}).}
\label{table1}
\end{table}
Unless otherwise stated we will employ the conventions $\alpha^{\prime}=2$
for the fermionic string and $\alpha^{\prime}=1$ for the bosonic string.

\section{The Ground Ring}
The
spin zero ghost number zero BRST invariant operators generate a commutative,
 associative
ring
\beq
{\cal O}(z){\cal O}'(0)\sim {\cal O}''(0)+\{Q,\dots\} \ ,
\label{BRS}
\eeq
called the ground ring, where $Q$ is the BRST operator.
Recall that for the $c=1$ string at the self-dual radius \cite{Lian:gk},
the ground ring is generated by four elements $x_{ij}, i,j=1,2$
with the relation \cite{Witten:1991zd}
\beq
x_{11}x_{22}-x_{12}x_{21} = 0 \ .
\label{con}
\eeq
Note that $x_{11},x_{22}$ are momentum states, while $x_{12},x_{21}$ are
winding states. 
Viewed as a complex surface, equation (\ref{con}) describes the conifold.

The ring structure (\ref{con}), which after adding the cosmological constant
$\mu$ was argued
to get deformed 
to 
 \beq
x_{11}x_{22}-x_{12}x_{21} = \mu \ ,
\label{cond}
\eeq   
was a first hint that the  $c=1$ string at the self-dual
radius is equivalent 
to a topological model on the deformed conifold geometry (\ref{cond})
 \cite{Ghoshal:1995wm}
\footnote{For the topological Landau-Ginzburg description, see \cite{Ghoshal:1993qt,Hanany:1994fi}.}.

In the following we will analyse the ground ring for the circle
line theories and the super affine theories at the radii (\ref{topp}).
We will consider first the theories  at $\mu=0$, and discuss
the deformation $\mu\neq 0$ later. The relevant BRST analysis 
has been performed in  \cite{Itoh:1991ix,Bouwknegt:1991va,Kutasov:1991qx,Bouwknegt:1991am}.

Consider the left sector.
The chiral BRST cohomology of dimension zero and ghost number zero is given by the 
infinite set of states
$\Psi_{(r,s)}$ with $r,s$ negative integers
\beq
\Psi_{(r,s)}\sim O_{r,s} e^{\left(ik_{r,s} x_L-p_{r,s}\Phi_L\right)} \ .
\eeq
The Liouville and matter momentum are given by 
\beq
k_{r,s}=\frac{1}{2}(r-s),\quad p_{r,s}=\frac{1}{2}(r+s+2) \ .
\eeq
The operators $\Psi_{(r,s)}$  are in the NS-sector if 
$k_{r,s}=(r-s)/2$ takes integer values, and in the R-sector if it takes half integer values. 

Of particular relevance for us are the R-sector operators:
\begin{eqnarray}\label{chiralring}
x(z)&\!\equiv\!&\Psi_{(-1,-2)}(z) 
\!=\!
\left( e^{-{i\over 2}H}e^{-{1\over 2}\vphi}
-{1\over\sqrt{2}}e^{{i\over 2}H}\partial\xi e^{-{3\over 2}\vphi}
\right) e^{{i\over 2}x-{1\over 2}\phi} \ , \nn
y(z)&\!\equiv\!&\Psi_{(-2,-1)}(z) 
\!=\!
\left( 
e^{{i\over 2}H} e^{-{1\over 2}\vphi}
-\frac{1}{\sqrt{2}}e^{-{i\over 2}H}\partial\xi e^{-{3\over 2}\vphi}
\right) e^{-{i\over 2}x-{1\over 2}\phi} \ ,
\eea
and the NS-sector operators 
\beq
\label{NSCP1}
u(z)\equiv\Psi_{(-1,-3)}(z)=x^2,~~~ 
v(z)\equiv\Psi_{(-3,-1)}(z)=y^2,~~~ 
w(z)\equiv\Psi_{(-2,-2)}(z)=xy \ ,
\eeq
given by 
\bea
u(z)
&=&
\left(-e^{-i H}e^{-\vphi}
+{i\over\sqrt{2}}c\partial\xi\partial(x-i\phi)e^{-2\vphi} %\nn
-\sqrt{2}c(\partial^2\xi-\partial\xi\partial\vphi)e^{-2\vphi}
\right)e^{ix-\phi} \ ,\
\nn
w(z)
&=&\left(
{2\sqrt{2}\over 3}i\partial H\,e^{-\vphi}
+{i\over 3}c\partial\xi
\left[\partial(x+i\phi)e^{-iH}-\partial(x-i\phi)e^{iH}\right]
e^{-2\vphi}\right. \nn 
&&\left.+c(\partial^2\xi-\partial\xi\partial\vphi)(e^{iH}+e^{-iH})
e^{-2\vphi}
-c\partial\xi\partial(e^{iH}+e^{-iH})e^{-2\vphi}\phantom{\frac{2}{2}}\right)e^{-\phi} \ ,\
\nn
v(z)
&=&
\left(-e^{i H}e^{-\vphi}
+{i\over\sqrt{2}}c\partial\xi\partial(x+i\phi)e^{-2\vphi} %\nn
+\sqrt{2}c(\partial^2\xi-\partial\xi\partial\vphi)e^{-2\vphi}
\right)e^{-ix-\phi} \ .
\end{eqnarray}

One has the multiplication rule
\begin{equation}
(\Psi_{(r,s)}\Psi_{(r^{\prime},s^{\prime})})(z)\sim
\Psi_{(r+r^{\prime}+1,s+s^{\prime}+1)}(z) \ ,
\end{equation}
where $\sim$ indicates that the right hand side could be multiplied by a vanishing constant.
The left sector ring of  spin zero, ghost number zero, and BRST invariant operators
is generated by the elements $x$ and $y$
\begin{eqnarray}\label{ringrel}
 \Psi_{(r,s)}=x^{-s-1}y^{-r-1},\quad\quad r,s\in{\mathbb Z}_- \ . 
\end{eqnarray}
Similarly, one can construct the ring in the right sector.

In order to construct the ground ring, we combine the left and right sectors
with the same left and right Liouville momenta.
Denote:
\begin{eqnarray}\label{ringelements}
  \begin{array}{ccc}
a_{ij}=\left(\begin{array}{cc}x\bar x&x\bar y\\
                         y\bar x&y\bar y \end{array}\right),
&\quad\quad&b_{ij}=\left(\begin{array}{ccc} u\bar u & u\bar w & u\bar v\\
                           w\bar u & w\bar w & w\bar v\\
                           v\bar u & v\bar w & v\bar v\end{array}\right) \ .
\end{array}
\end{eqnarray}
Clearly,
\beq
det(a_{ij}) = 0 \ ,
\label{det}
\eeq
which is the conifold equation (\ref{con}).
The relations among the $b_{ij}$ similarly follow from
(\ref{NSCP1}).
Note that 
$a_{ij}$ are in the RR sector while $b_{ij}$ are in the NS-NS sector.
Note also that   $a_{12},a_{21}$ are winding operators, $a_{11},a_{22}$ 
are momentum operators, 
$b_{13},b_{31}$ are winding operators and  $b_{11},b_{22},b_{33}$ momentum operators. 
$b_{12},b_{21},b_{23},b_{32}$ are mixed winding/momentum operators.

Next we need to impose the GSO projection.
We will consider the circle line theories and the superaffine theories at the radii (\ref{topp}).

\subsection{Circle Line Theories}

Since type 0A and type 0B are T-dual to each other under $R\rightarrow \frac{\alpha'}{R}$,
we essentially have to consider only one of them\footnote{For a discussion of T-duality
in finite flux backgrounds see e. g. \cite{Gross:03}}.
Consider type 0A at the radius  $R=\frac{l_s}{\sqrt{2}}=1$.
We impose GSO projection on the operators (\ref{ringelements}) (in the appropriate
momenta and winding lattices). In the NS-NS sector we have
\begin{eqnarray}
  mm:\quad k=n \in{\mathbb Z},\quad &&wm:\quad k=\frac{w}{2},w\in{\mathbb Z} \ ,
\end{eqnarray}
while in the RR sector only winding modes are allowed,
\begin{eqnarray}
  wm:\quad k=\frac{w}{2},w\in{\mathbb Z} \ .
\end{eqnarray}
The ground ring elements that survive the GSO projection are
\begin{eqnarray}
  a_{ij}=\left(\begin{array}{cc}&{\bf x\bar y}\\{\bf y\bar x}&\end{array}\right),\quad b_{kl}=\left(\begin{array}{ccc}{\bf u\bar u}&&u\bar v\\&w\bar w&\\v\bar u&&{\bf v\bar v}\end{array}\right) \ ,\label{0AR1}
\end{eqnarray}
where we denoted
by bold letters the generators of the ground ring. For instance, $b_{22}$ is generated as 
$a_{12}a_{21}=b_{22}$, whereas $b_{11}$ cannot be generated from the $a_{ij}$.
Thus, the ground ring at $\mu=0$ is generated by four elements   $a_{12}, a_{21}, b_{11},b_{33}$
with the relation
 \begin{eqnarray}
\label{GR0AB1} 
 (a_{12})^2(a_{21})^2-b_{11}b_{33}=0 \ .
\label{conz2}
\eea
The complex 3-fold (\ref{conz2}) is the $Z_2$ quotient of the conifold
(\ref{det}), with the $Z_2$ action being
\bea
Z_2:~~~~(a_{11},a_{22}) &\rightarrow& -(a_{11},a_{22}) \nn
(a_{12},a_{21}) &\rightarrow& (a_{12},a_{21}) \ .
\label{Z2}
\end{eqnarray}
We therefore suggest that 
{\it type 0A (0B) at the radius $R=1$ ($R=2$) is equivalent to the 
topological B-model on the (deformed) $Z_2$ quotient of the
conifold (\ref{Z2})}.

Note that, $c=1$ bosonic string theory at half the self-dual radius
is a 
$Z_2$ orbifold of the self-dual
radius case \cite{Ghoshal:1992kx}. 
The action on the ground ring elements $x_{ij}$ (\ref{con}) 
is (\ref{Z2}), and therefore the invariant ring 
is as the type 0A one (\ref{conz2}). 

\noindent
{\it Deformation of the ground ring}

When the cosmological constant $\mu$ and the background RR charge $q$
are nonzero\footnote{We will consider a nonzero RR charge of one RR field.}, 
one expects a deformation of the ground ring 
relation.
Consider first the case   $\mu\neq 0$ and $q=0$.
One possibility is to take the $Z_2$
quotient of the
deformed version of (\ref{det})
\beq
det(a_{ij}) = \mu \ .
\eeq
This leads to
\beq
(a_{12}a_{21} + \mu)^2  -b_{11}b_{33}=0 \ .
\label{s22}
\eeq
Note, that (\ref{s22})
is singular at 
\beq
a_{12}a_{21} + \mu = 0, ~~b_{11}=b_{33}=0 \ .
\label{gr0A}
\eeq

Matching the Liouville momentum between the deformation and the ground ring 
relation (\ref{conz2}) \cite{Ghoshal:1993ur}, allows
another deformed ring relation
\beq
(a_{12}a_{21} +\mu)(a_{12}a_{21}-\mu) -b_{11}b_{33}=0 \ .
\label{s1}
\eeq
Note that the complex hypersurface described by (\ref{s1}) is non-singular.
However, we will argue in section 6 that the deformed
ring relation (\ref{s22}) is the expected one.

Consider next the case $q\neq 0$. Since the theory
has the discrete symmetry $q \to -q$, we expect only even powers of $q$ to appear.
Moreover, the  Liouville momentum considerations suggest that only a $q^2$
term will appear in the deformed ring relation.
We will argue in section 6, that the expected deformation is
\beq
\label{Geomy11}
(a_{12}a_{21} +\mu)^2= b_{11}b_{33} -\frac{q^2}{4} \ .
\end{equation}

One can identify two 3-spheres in the geometry (\ref{Geomy11}) as follows.
First introduce a complex parameter $t$ defined by
\begin{equation}
a_{12}a_{21}=t \, .
\end{equation}
Equation (\ref{Geomy11}) can be rewritten as
\begin{equation}
b_{11}b_{33}=(t + \mu +\frac{i q}{2})(t + \mu -\frac{i q}{2}) \  .
\end{equation}

We now regard the geometry of  (\ref{Geomy11}) as a $\bC\times\bC$ fibration
over the complex plane spanned by $t$, where the coordinates of
the first and second fibration are given by $a_{12},a_{21}$
and $b_{11},b_{33}$, respectively.
Each fibration has a topology of the product of two cylinders, whose
closed cycles are generated by 
$(a_{12},a_{21})\rightarrow (e^{i\theta_1}a_{12},e^{-i\theta_1}a_{21})$
and
$(b_{11},b_{33})\rightarrow (e^{i\theta_2}b_{11},e^{-i\theta_2}b_{33})$.

At a generic point of $t$, the fibration is regular. There are three
special points where the fibers get degenerate:
at $t=0$, the first fibration develops a degenerate $\bS^1$, while the
second is smooth.
At $t=-\mu \pm \frac{i q}{2}$, the second fibration degenerates, while the first is
regular. To see the two $\bS^3$'s, consider
%the following
\begin{equation}
C_{\pm}:~~I_{\pm}\times \bS^1_1\times\bS_2^1 \, .
\end{equation}
Here $I_{\pm}$ are the interval on the complex plane between $t=0$ and
$t=-\mu \pm \frac{i q}{2}$. $\bS^1_{1,2}$ denote the closed cycles of the first and
second cylinders over the intervals.
Thus, $C_{\pm}$ define two $\bS^3$'s.
Therefore, compared to the deformed conifold analysis \cite{Strominger:1995cz},
we expect two rather than one BPS hypermultiplet from D-brane wrappings in type II compactification.
Using \cite{Vafa:1995ta}
\beq
F_1 = -\frac{1}{12}\sum \ln~ m_{BPS}^2 \ ,
\label{BPS}
\eeq
we get at $q=0$\footnote{We will discuss the case $q\neq 0$ as well as the higher genus
partition functions $F_g$ in section 6.}
\beq
F_1=-2{1\over 12}\ln\mu = -\frac{1}{6}\ln\mu  \ ,
\eeq
which agrees with 
$F_{0A}(\sqrt{\alpha'/2})$.

Note in comparison, that for
$c=1$ bosonic string theory at half the self-dual radius, the
one-loop partition function of this theory computed in \cite{GK:90}
\beq
F_1=-{1\over 24}\left(2+{1\over 2}\right)\ln\mu\ ,
\eeq
is not an integer multiple of $-{1\over 12}\ln\mu$.

\subsection{Super Affine Theories}

The super affine theories are self-dual at the radius 
$R_{\rm super-affine}=l_s\sqrt{2}=2$. We consider the ground ring at this radius.

\begin{itemize}

\item{} Super affine 0A:  The GSO projection acts on the
elements of the ground ring (\ref{ringelements}) by $a_{ij}\rightarrow -a_{ij}$
keeping $b_{ij}$ invariant.
Therefore the RR operators are projected out, while the NS-NS states remain with
\begin{eqnarray}
  mm:\quad k=n \in{\mathbb Z},\quad &&wm:\quad k=w\in{\mathbb Z} \ .
\end{eqnarray}
The ground ring is generated by
the invariant elements $b_{ij}$ subject to the conditions (\ref{det}) and the projection 
 $a_{ij}\rightarrow -a_{ij}$. 
The complex 3-fold described by the ground ring  is the $Z_2$ quotient of the conifold
(\ref{det})
\beq
Z_2:~~~~a_{ij} \rightarrow -a_{ij} \ .
\label{z22}
\eeq
We therefore suggest that 
{\it super affine  0A at the radius $R=2$ is equivalent to the
topological B-model on the (deformed) $Z_2$ quotient of the
conifold (\ref{z22})}.

\item{} Super affine 0B:  The GSO projection keeps in (\ref{ringelements})
all the RR operators $a_{ij}$ with
\begin{eqnarray}
  mm:\quad k=\frac{n}{2}~~~ n\in{\mathbb Z},\quad &&wm:\quad k= \frac{w}{2}~~~
w\in{\mathbb Z} \ ,
\end{eqnarray}
and all the
NS-NS operators ($b_{ij}$) with 
\begin{eqnarray}
  mm:\quad k=n \in{\mathbb Z},\quad &&wm:\quad k=\frac{w}{2},~~~w\in{\mathbb Z} \ .
\end{eqnarray}
The ground ring is generated by
the invariant elements $a_{ij}$ subject to the conditions (\ref{det}).
The complex 3-fold described by the ground ring  is the conifold
(\ref{con}).
We therefore suggest that 
{\it super affine  0B at the radius $R=2$ is equivalent to  the
topological B-model on the (deformed)
conifold}.
\end{itemize}

\noindent
{\it Deformation of the ground ring}\\

The singular geometry described by the super affine 0B ground ring at $\mu=0$
is the
conifold (\ref{con}). 
When $\mu \neq 0$ we expect the ground ring to get deformed in the unique
$SU(2)\times SU(2)$ invariant way to (\ref{cond}).

Recall that the $c=1$ bosonic string at the self-dual radius is also described by  the
topological B-model on the deformed
conifold. This  implies that the two theories are equivalent {\it perturbatively}.
In particular we should have
perturbatively 
\bea
\cF^{\rm super-aff}_{\rm 0B}(R_{\rm self-dual})={1\over 2}\mu^2 \log\mu -{1\over 12} \log\mu + \nn
{1\over 240}\mu^{-2}+\sum_{g>2} a_g \mu^{2-2g}  
=  \cF_{c=1}(R_{\rm self-dual}) \ .
\label{part}
\eea
$a_g = \frac{B_{2g}}{2g(2g-2)}$ is the Euler class of 
the moduli space of Riemann surfaces of genus $g$. 
Indeed, at genus one, 
$F_{\rm 0B}^{\rm super-aff.}(R_{\rm self-dual})= F_{c=1}(R_{\rm self-dual})=
-\frac{1}{12}\ln\mu$.

We expect the super affine  0B at the radius $R=2$ to provide a {\it non-perturbative} 
completion of topological B-model on the deformed
conifold.

The singular geometry described by the ground ring
of super affine 0A theory at the self-dual
radius and $\mu=0$
is that of a Calabi-Yau space, where a three cycle of the form $S^3/Z_2$
shrinks to zero size.
When $\mu\neq0$ we expect a deformation of the space such that
$Vol(S^3/Z_2) \sim \mu/2$. The Calabi-Yau looks locally like $T^*(S^3/Z_2)$.
This is the unique $SU(2)\times SU(2)$ invariant deformation (see section 7).

When a three cycle $S^3/\Gamma$ shrinks to zero size in a Calabi-Yau space, 
with $\Gamma$ a freely acting
discrete subgroup of $SU(2)$, then in the deformed
background one expects 
\beq
F_1 = -\frac{|\Gamma|}{12} \ln \mu \ .
\label{G}
\eeq
$|\Gamma|$ is the order of the group, and (\ref{G}) is derived by
counting the number of BPS D-brane wrapping states \cite{Gopakumar:1997dv}
and using (\ref{BPS}).
When $\Gamma = Z_2$ we get $F_1= -\frac{1}{6}\ln\mu$, which is in agreement
with 
$F_{\rm 0A}^{\rm super-aff.}(R_{\rm self-dual})$.

In the remainder of the paper we will study some topological and integrable aspects
of $\hat{c}=1$ strings. We will analyse partition functions, Ward 
 identities, correlators, the topological B-model and the relation to quiver gauge theories.

\section{Matrix Models}

In this section we will analyse the matrix models of type 0A and 
superaffine 0A theories
on a circle. The type 0A partition function has been computed in
\cite{Douglas:2003up}. In this section we will compute
the superaffine 0A partition function.
We will find that {\it non-perturbatively} at the special radii (\ref{topp})
\beq
\cF^{\rm super-aff.}_{\rm 0A}(R=\sqrt{2\alpha'}) = \cF_{\rm 0A}(R=\sqrt{\alpha'/2}) \ ,
\eeq
when the RR charge $q=0$. As we will see, 
this equality seems to work also for the momentum states
tachyon correlators, which suggests
that superaffine 0A at the self-dual radius describes
a $(q=0)$ NS-NS sector of 0A at radius $\sqrt{\alpha'/2}$.
We will see that
{\it perturbatively}
\beq
\cF^{\rm super-aff}_{\rm 0A}(R_{\rm self-dual})=
2\cF_{c=1}(R_{\rm self-dual}) \ ,
\label{part2}
\eeq
and of course the same for $\cF_{\rm 0A}(\sqrt{\alpha'/2})$.
We will see later that the relations (\ref{part}) and (\ref{part2}) appear 
in other descriptions of the systems
such as the quiver $\hat{A}_1$ matrix model and its $Z_2$ quotients.

These suggest that the (perturbative) partition function of the topological B-models
on the $Z_2$ quotients of the conifold 
(\ref{Z2}) and (\ref{z22}), as a function of the deformation parameter,
is twice that of the 
$c=1$ bosonic string  at the self-dual radius.
This is indeed expected from topological considerations \cite{Gopakumar:1998vy}, namely
\beq
\cF_{S^3/Z_2}(\mu) = 2 \cF_{S^3}(\frac{\mu}{2}) \ ,
\eeq
where  $\cF_{S^3}$ is the partition function of the 
topological B-model on the (deformed) conifold 
and $\cF_{S^3/Z_2}$ is the partition function of the 
topological B-model on the (deformed) $Z_2$ quotient of the conifold.
This follows from (\ref{part2}) with 
the conventions such that $R_{\rm self-dual}=1$ in both theories.

\subsection{The Matrix Model Analysis}

The type 0A matrix model
is the theory on the worldvolume
of $N+q$ $D0$-branes and $N$ $\bar{D0}$-branes of type 0A string theory.
It is a
$ U(N) \times U(N+q) $ theory with a tachyon
field described by a complex matrix $t(\tau)$ in the bifundamental \cite{Douglas:2003up}.

The superaffine 0A matrix model on a circle $R$ is obtained  
from type 0A theory compactified on a circle with the radius
$\cR$ by 
imposing  the boundary condition \cite{Douglas:2003up}
\begin{equation}
t(\tau+2\pi \cR)=t^{\dagger}(\tau) \ ,
\label{bc;saffine}
\end{equation}
with $\cR={R\over 2}$.
To see this, recall that the superaffine theory is defined by taking
the $\bZ_2$ quotient $(-1)^{{\bf F}_L}e^{i\pi p}$, 
where $p\in\bZ$ is momentum $k={p\over \cR}$ and ${{\bf F}_L}$ is
left space-time fermion number.
In the NS-NS sector, the quotient is implemented by using 
$\cR={R\over 2}$, which implies even integer momentum.
In the RR sector, the RR one-form potential is odd under the
$\bZ_2$ action, which exchanges the gauge groups on the D-branes worldvolume
$U(N+q) \leftrightarrow U(N)$.
This is a symmetry only for $q=0$, and 
is consistent with the fact that the physical spectrum of the
superaffine 0A theory does not include RR states.

We expand $t$ as
\begin{equation}
t(\tau)=\sum_n t_n e^{in\tau/(2\cR)} \ .
\end{equation}
It then follows from (\ref{bc;saffine}) that $t_n$ obey
\begin{equation}
t_n^{\dagger}=(-1)^n t_{-n} \ .
\end{equation}
Decompose now $t(\tau)$ as 
\beq
t(\tau)={1\over\sqrt{2}}(t_+(\tau)+t_-(\tau)) \ ,
\label{deco}
\eeq
 where
\begin{equation}
t_+(\tau)=\sum_{n={\rm even}}t_n e^{in\tau/(2\cR)} , \ ~~
t_-(\tau)=\sum_{n={\rm odd}}t_n e^{in\tau/(2\cR)} \ .
\end{equation}
One finds that
\begin{equation}
\left(t_+(\tau)\right)^{\dagger}=t_+(\tau) \ , ~~
\left(t_-(\tau)\right)^{\dagger}=-t_-(\tau) \ .
\end{equation}
Hence the matrix model is given by two decoupled (anti-)Hermitian
matrices
\begin{eqnarray}
S&\!=\!&\beta\int_0^{2\pi \cR}d\tau\, \tr\left(
\dot{t}^{\dagger}\dot{t}+{1\over 2\alpha^{\prime}}t^{\dagger}t
\right) \nn
&\!=\!&
{\beta\over 2}\int_0^{2\pi \cR}d\tau \,\,\tr\left(
\dot{t}_+^2+{1\over 2\alpha^{\prime}}t_+^2
-\dot{t}_-^2-{1\over 2\alpha^{\prime}}t_-^2
\right) \ .
\label{maction:sa0a}
\end{eqnarray}
Note that
\begin{equation}
t_+(\tau+2\pi \cR)=t_+(\tau) \ ,~~
t_-(\tau+2\pi \cR)=-t_-(\tau) \ .
\end{equation}

We note that the complex matrix model (\ref{maction:sa0a}) is invariant
under the $U(N)_L\times U(N)_R$ group acting by $t\rightarrow U_L\,t\,U_R^{\dagger}$.
With the decomposition of $t$ (\ref{deco}), the manifest symmetry is 
$U(N)_V\times U(1)_A$, where
$t_{\pm}$ transform in adjoint representation of $U(N)_V$, and are in a
doublet of $U(1)_A=O(2)_A$.
Since the matrix model is gauged, the Gauss law constraint
implies that the physical states are invariant under 
$U(N)_V\times U(1)_A$.
To see how the constraint is implemented, consider for simplicity
the case $N=1$. 
$t_{\pm}$ yield two independent harmonic oscillators with the creation
operators given by $a_{\pm}^{\dagger}$.
They transform in a doublet of $U(1)_A$ and are invariant under
$U(1)_V$. It thus follows from the Gauss law constraint that the
physical states are generated by gauge invariant operators. 
This corresponds to the doubling of energies observed in the 0A matrix model
\cite{Douglas:2003up}.

\subsection{A Brief Review}

In the following we will briefly review some basic aspects
of matrix models, which 
will be needed for the computation of the super affine 0A partition function
(For a review, see for instance \cite{review:c=1}).
Consider a Hermitian matrix model with the action given by
\begin{equation}
S=\int_0^{2\pi R}d\tau\, \beta\,\tr \left(
\dot{X}^2+{1\over 2\alpha^{\prime}}X^2\right) \ ,
\end{equation}
and with the boundary condition 
\beq
X(\tau+2\pi R)=\pm X(\tau) \ .
\eeq
The corresponding Hamiltonian reads:
\begin{equation}
H=\beta\,\tr\left( {1\over 2\beta}p^2-{1\over 4\alpha^{\prime}}X^2\right)
\equiv \beta h \ .
\end{equation}
We decompose the matrix X as
\begin{equation}
X=U\left(
\begin{array}{ccc}
\lambda_1 && \\
& \ddots & \\
&& \lambda_N 
\end{array}
\right) U^{\dagger} \ ,~~~ U\in U(N) \ .
\end{equation}
Under this decomposition, the path integral measure reads
\begin{equation}
DX=\prod_{i=1}^N d\lambda_i d\Omega_{U(N)}\Delta(\lambda)^2 \ ,
\end{equation}
where
\begin{equation}
\Delta(\lambda)=\prod_{i<j}(\lambda_i-\lambda_j) \ .
\end{equation}
The angular part $d\Omega_{U(N)}$ is not relevant for our purposes
and will be omitted in the discussion.
Upon quantization, the Hamiltonian reads
\begin{eqnarray}
\hat{h}&\!=\!&\tr\left(
-{1\over 2\beta^2}{\partial^2\over \partial X^2}
-{1\over 4\alpha^{\prime}}X^2\right) \nn
&\!=\!&\sum_{i=1}^N\left(
-{1\over 2\beta^2}{1\over\Delta}{\partial^2\over\partial\lambda_i^2}\Delta
-{1\over 4\alpha^{\prime}}\lambda_i^2\right) \ .
\end{eqnarray}
The Schr\"odinger equation is
\begin{equation}
\hat{h}\Psi(\lambda)=e\Psi(\lambda) \ .
\end{equation}
Defining the wave function $\psi=\Delta\Psi$,
the Schr\"odinger equation reads
\begin{equation}
\sum_{i=1}^N\left[
-{1\over 2\beta^2}{\partial^2\over\partial\lambda_i^2}
-{1\over 4\alpha^{\prime}}\lambda_i^2\right]\psi=e\psi \ .
\end{equation}
Thus,
$\psi$ 
describes a system of $N$ free fermions.

Consider the partition function of the matrix model
\begin{equation}
Z=\int DX\,e^{-S} \ .
\end{equation}
It is easy to show that this is equivalent to
\begin{equation}
Z_{\Omega}=\tr\left( \Omeh\, e^{-2\pi R\beta\hat{h}}\right) \ ,
\end{equation}
where $\Omeh=1$ for the periodic $X$. For the anti-periodic $X$,
$\Omeh$ is a $\bZ_2$ operation defined by
\begin{equation}
\Omeh|X\rangle = |-X\rangle \ ,
\end{equation}
where $|X\rangle$ is an eigenstate of the Schr\"odinger operator
$\hat{X}$.
$Z_{\Omega}$ can be recast as
\begin{equation}
Z_{\Omega}=\int de\,e^{-2\pi R\beta e}\,\rho_{\Omega}(e) \ ,
\end{equation}
where $\rho_{\Omega}(e)$ is the ``twisted'' density of states defined by
\begin{equation}
\rho_{\Omega}(e)=\tr\left( \Omeh\,\delta(\hat{h}-e)\right) \ .
\end{equation}
This can be written as
\begin{equation}
\rho_{\Omega}(e)={\beta^2\over \pi}{\rm Im}
\int_{-\infty}^{\infty}d\lambda\,
\langle\lambda|\Omeh
{1\over -{1\over 2}{\partial^2\over\partial\lambda^2}
        +{1\over 2}\beta^2\omega^2\lambda^2-\beta^2e-i\epsilon}
|\lambda\rangle \ ,
\end{equation}
with $\omega^2=-1/(2{\alpha^{\prime}})$. 
Note that the integration range of $\lambda$ is
$-\infty<\lambda<\infty$. 
This reflects the fact that both sides of 
the potential are filled with fermions symmetrically, 
as in the type 0B matrix model.
Using 
\begin{eqnarray}
&&\langle\lambda_f|
{1\over -{1\over 2}{\partial^2\over\partial\lambda^2}
        +{1\over 2}\omega^2\lambda^2+e-i\epsilon}
|\lambda_i\rangle \nn
&&=
\int_0^{\infty}dt e^{-et}\sqrt{\omega\over 2\pi\sinh(\omega t)}\,
e^{-\omega\left((\lambda_i^2+\lambda^2_f)\cosh(\omega t)
                -2\lambda_i\lambda_f\right)/(2\sinh(\omega t))} \ ,
\end{eqnarray}
one gets for $\Omeh=1$
\begin{equation}
\rho_+(e)={\beta^2\over\pi}{\rm Re}\int_0^{\infty}dt\,
{e^{i\beta^2 et}\over 2\sinh\left(\beta t\over 2\sqrt{\alpha}\right)} \ ,
\end{equation}
while for the anti-periodic case 
\begin{equation}
\rho_-(e)={\beta^2\over\pi}{\rm Re}\int_0^{\infty}dt\,
{e^{i\beta^2 et}\over 2\cosh\left(\beta t\over 2\sqrt{\alpha}\right)} \ .
\end{equation}
Here $\alpha=2\alpha^{\prime}$.

In order to evaluate the partition function, it is convenient to 
introduce a chemical potential such that \cite{GK:90}
\begin{equation}
N=\int de \,\rho(e){1\over 1+e^{2\pi R\beta(e-\mu_F)}} \ .
\end{equation}
Define
\begin{eqnarray}
&&g={N\over\beta} \, ~~~\Delta=1-g \ , \nn
&&\mu=\mu_c-\mu_F \ , \nn
&&x=\mu_c-e \ ,
\end{eqnarray}
where $\mu_c$ is the critical point defined by 
$\mu_c=V(\lambda_c),~V^{\prime}(\lambda_c)=0$. 
In the our  case, $\mu_c=0$.
The double scaling limit reads
\begin{equation}
\beta\sim N\rightarrow \infty, ~~ \beta\mu={\rm finite} \sim {1\over g_s} \ .
\end{equation}

Note that the free energy $F$ is a function of $N$ satisfying
\begin{equation}
{1\over 2\pi R\beta}{\partial F\over \partial N}=\mu_F \ .
\end{equation}
To get the free energy as a function of $\mu=-\mu_F$, one performs
a Legendre transform \cite{GK:90}:
\begin{equation}
-\cF\equiv F-2\pi R\beta N\mu_F \ .
\end{equation}
One can verify that $\cF$ obeys
\begin{equation}
{\partial^2\cF\over \partial\mu^2}=2\pi R\beta^2
{\partial\Delta\over\partial\mu} \ ,
\end{equation}
and $\cF$ can be obtained by integrating this equation.

\subsection{Solution of Super Affine $0$A Matrix Model}

Consider now the matrix model (\ref{maction:sa0a}).
We will first evaluate the density of states
\begin{equation}
\rho_{\rm S0A}(e)=\tr (\Omeh\,\delta(\hat{h}-e)) \ ,
\end{equation}
where the trace is taken over all the physical states of the two harmonic 
oscillators obeying the Gauss law constraint. $\Omeh$ acts only on the
Hilbert space of $t_-$ as $\Omeh |t_-\rangle = |-t_-\rangle$,
in order to implement the anti-periodic boundary condition for $t_-$.
$\Omeh$ acts on the oscillators as
\beq
\Omeh (a_-,a_-^{\dagger})\Omeh=-(a_-,a_-^{\dagger}) \ .
\eeq
To see this, recall that $\Omeh$ acts on the operator $\hat{t}_-$ as
$\Omeh\, \hat{t}_-\Omeh=-\hat{t}_-$, which is consistent with the
relation $\hat{t}_-=(a_-+a_-^{\dagger})/\sqrt{2}$.
As discussed before, the physical states 
%take the form
%\begin{equation}
%|n\rangle \sim \left((a_+^{\dagger})^2+(a_-^{\dagger})^2\right)^n |0\rangle \ .
%\end{equation}
are in a Hilbert space of a single harmonic
oscillator with a doubling of the energy
\begin{equation}
\hat{h}|n\rangle =2e_n|n\rangle \ ,
\end{equation}
$e_n$ being the energy of a single harmonic oscillator with the
occupation number $n$.
Note that
the physical state $|n\rangle$ is invariant
under $\Omeh$. 
Thus, the computation of the density of states of superaffine 0A matrix
model is equivalent to that of a single Hermitian matrix with a 
doubling of the energy:
\begin{eqnarray}
\rho_{\rm S0A}(e)&\!=\!&\sum_n\delta (2e_n-e)
={1\over 2}\sum_n\delta(e_n-{e\over 2}) \nn
&\!=\!&
{\beta^2\over 4\pi}{\rm Re}\int_0^{\infty}dt\,
{e^{{i\over 2}\beta^2 et}\over 
2\sinh\left(\beta t\over 2\sqrt{\alpha}\right)} \ .
\end{eqnarray}
This  is exactly the result for type 0A matrix model when $q=0$ \cite{Douglas:2003up}.

Note that the above computation was done by filling both sides of an upside
down potential with a symmetric Fermi level, and
as we will see, it agrees with the worldsheet one-loop
partition computed in \cite{Douglas:2003up}.
The symmetric Fermi level ensures that the superaffine 0A theory is
stable nonperturbatively.

The boundary condition (\ref{bc;saffine}) means that the only
difference between the 0A and superaffine 0A on a circle is the radius
of the circle. The partition function of the superaffine 0A
theory on a circle with radius $R$ is identical to the partition function of type 0A 
on a circle with radius $\cR={R\over 2}$.
Following the procedures reviewed before, we obtain
\begin{equation}
{\partial^2\cF_{\rm S0A}\over \partial\mu^2}
=\pi R\beta^2{\partial\Delta_{\rm S0A}\over\partial\mu} \ ,
\end{equation}
with
\begin{equation}
{\partial\Delta_{\rm S0A}\over\partial\mu}=
{\sqrt{\alpha^{\prime}/2}\over\pi\mu}\,{\rm Im}\int_0^{\infty}dt\,e^{-it}
\,
{t/(2\beta\mu\sqrt{\alpha^{\prime}/2})\over
\sinh(t/(2\beta\mu\sqrt{\alpha^{\prime}/2}))}\,
{t/(\beta\mu R)\over\sinh(t/(\beta\mu R))} \ .
\end{equation}
This is invariant under the self-duality transformation \cite{Douglas:2003up}
\begin{equation}
R\rightarrow {2\alpha^{\prime}\over R} \ ,~~~
\beta\mu\rightarrow {R\over \sqrt{2\alpha^{\prime}}} \beta\mu \ .
\label{seldual:soa}
\end{equation}

An alternative simple way to solve the superaffine 0A matrix model
is to decompose the $N\times N$ complex matrix as \cite{cpxmatrix}
\begin{equation}
t=V\left(
\begin{array}{ccc}
\lambda_1 && \\
& \ddots & \\
&& \lambda_N 
\end{array}
\right) W^{\dagger} \ ,~~~ V,W\in U(N) \ .
\end{equation}
Up to the angular part the path integral measure reads
\begin{equation}
\prod_i d\lambda_i\lambda_i\,\prod_{i<j}(\lambda_i^2-\lambda_j^2)^2 
\equiv \prod_i d\lambda_i\,J \ ,
\label{jacobian}
\end{equation}
at each time of $\tau$.
The boundary condition (\ref{bc;saffine}) implies
\begin{equation}
\lambda_i(\tau+2\pi\cR)=\lambda_i(\tau) \ ,~~
V(\tau+2\pi\cR)=W(\tau) \ .
\label{VW}
\end{equation}
The path integral with respect to the angular part gives
a volume factor of the gauge group, and we will see that we end up with the deformed
matrix model \cite{Jevicki:1993zg}, with $q=0$ and the compactification radius given by $\cR$.

It follows from (\ref{jacobian}) that upon quantization,
the Schr\"odinger equation of the superaffine 0A matrix model takes the
form
\begin{equation}
\sum_{i=1}^N\left[
-{1\over 2J}{\partial\over\partial\lambda_i}
\left( J{\partial\over\partial\lambda_i}\right)
-{1\over 4\alpha^{\prime}}\lambda_i^2\right]\Psi=E\Psi \ .
\end{equation}
Defining the wavefunction $\chi$ as
\begin{equation}
\Psi={1\over\sqrt{J}}\,\chi \ ,
\end{equation}
the Schr\"odinger equation can be written  as
\begin{equation}
\sum_{i=1}^N\left[
-{1\over 2}{\partial^2\over\partial\lambda_i^2}
-{1\over 4\alpha^{\prime}}\lambda_i^2
+{M\over 2\lambda_i^2}\right]\chi=E\chi \ ,
\end{equation}
with $M=-{1\over 4}$.
This is the Hamiltonian of the deformed matrix model
\cite{Jevicki:1993zg}. It
is described by a single Hermitian matrix $\Phi$ whose eigenvalue
is equal to $\lambda_i$ with the potential given by
\begin{equation}
V(\Phi)=\tr\left(-{1\over 4\alpha^{\prime}}\Phi^2+\frac{M}{2 \Phi^2} 
\right) \, .
\end{equation}

Consider ${\partial\Delta_{\rm S0A}\over\partial\mu}$ in detail.
One finds that 
\begin{eqnarray}
{\partial\Delta_{\rm S0A}\over\partial\mu}
&\!=\!&
{\sqrt{\alpha^{\prime}/2}\over\pi}\left[
-\log\mu+\sum_{g\ge 1}
\,(\beta\mu\sqrt{\alpha})^{-2g}\,f_g(R)\right] \ ,
\end{eqnarray}
where
\begin{eqnarray}
f_g(R)=(2g-1)!\,
\sum_{n=0}^g
{(2^{2n}-2)(2^{2(g-n)}-2)\over (2n)!\,(2(g-n))!}
B_nB_{g-n}
\left({R\over \sqrt{\alpha}}\right)^{-2n} \ .
\end{eqnarray}
As an example,
\begin{equation}
f_1(R)={1\over 6}\left( 1+{\alpha\over R^2}\right) \ , ~~~
f_2(R)={1\over 6}\left( {7\over 10}+{\alpha\over R^2}
                       +{7\over 10}{\alpha^2\over R^4}\right) \ .
\end{equation}
It thus follows that
\begin{equation}
\cF_{\rm S0A}={R\over 4\sqrt{\alpha}}\left[
-\,(\beta\mu\sqrt{\alpha})^2\,\log\mu
-2\log\mu\,f_1(R)
+\sum_{g\ge 2}{(\beta\mu\sqrt{\alpha})^{2-2g}\over (g-1)(2g-1)}
\,f_g(R)\right] = \cF_{0A}(\cR) \ .
\end{equation}
The one-loop partition function agrees with the result of \cite{Douglas:2003up}.

Setting $R=\sqrt{\alpha}$, the self-dual radius, 
one finds that
\begin{equation}
\cF_{\rm S0A}=2\left[
-{1\over 2}(\beta\mu
%\sqrt{\alpha^{\prime}/2}
)^2\log\mu
-{1\over 12}\log\mu
+\sum_{g\ge 2}{B_{g}\over 2g(2g-2)}\,(\beta\mu
%\sqrt{\alpha^{\prime}/2}
)^{2-2g}
\right]= 2\cF_{c=1} \ ,
\end{equation}
being twice the partition function of bosonic $c=1$ string theory
at the self-dual radius.

\subsection{Tachyon Scattering}

In the following we will argue that the momentum mode tachyon correlators in  superaffine 0A
at radius $R$ are identical to the  momentum mode tachyon correlators in type 0A ($q=0$)
at radius $\cR={R\over 2}$.
With such relation between the radii there is a one to one map between the tachyon momentum states
of the two theories (see section 2).

We consider the tachyon scattering in the
matrix model framework.
As discussed before, the 0A matrix model is the theory of a gauged complex $N\times N$ matrix
$t$ with periodic boundary condition 
$t(\tau+2\pi\cR)=t(\tau)$.
In order to analyse the tachyon scattering we need first
to construct the corresponding operator in the
complex matrix model.
Here it is convenient to use the equivalence between the 0A matrix model and
the deformed matrix model.
The tachyon operator with a momentum $n$ in the
deformed matrix model is given by \cite{Demeterfi:1993sj} (see also appendix A)
\begin{equation}
T_n \sim \int d\tau\,e^{in\tau}\,\tr e^{-l\Phi^2} \ .
\end{equation}
This equals 
\begin{equation}
T_n \sim \int d\tau\,e^{in\tau}\,\tr e^{-l\,tt^{\dagger}} \ ,
\end{equation}
since $\Phi,t,t^{\dagger}$ have the same eigenvalues.

Consider next the momentum mode tachyons in  superaffine 0A.
We need to use the relation between the radii 
\beq
\cR={R\over 2} \ ,
\label{rad}
\eeq
and impose
the boundary condition $t(\tau+\pi R )=t^{\dagger}(\tau)$.
As discussed before, this modifies the boundary condition of the angular
part (\ref{VW}), $V(\tau+\pi R)=W(\tau)$, leaving the eigenvalues periodic.
Since the tachyon operator $T_n$ is gauge invariant,
this modification is irrelevant for it.
Therefore the tachyon operators as well as the matrix model action and gauge invariant
measure are the same for type 0A  matrix model and the  superaffine 0A matrix model
at radii (\ref{rad}).
Thus, we expect the same scattering amplitudes of 
the momentum mode tachyons in both theories.

\section{Integrable Structure}

The integrable structure of the $c=1$ bosonic string compactified
on a circle has been analysed in \cite{Dijkgraaf:1992hk}.
It was found that the Euclidean 2D string, perturbed purely by
tachyon momentum modes (or by winding modes in the T-dual picture),
has the integrable structure of the Toda lattice hierarchy.
More precisely, the generating functional of
the tachyon momentum modes correlators is a tau function $\tau(t, \bar{t})$
of the Toda lattice hierarchy, with the Toda times $\bar{t},t$ being 
the sources of the positive and negative momentum modes, respectively. 

The Toda hierarchy is quite general  and one has to provide additional
information in order to select the correct tau function, which is
specified by the so called string equations (see e.g.
\cite{Kostov:2001wv} in  the $c=1$ context). In other words,
one has to provide the boundary conditions for the Toda differential equations,
which determine the flows in the Toda times $t,\bar{t}$. In the
context of non-critical  string theory the boundary condition is determined by
the unperturbed partition function (see e.g. \cite{Kazakov:2000pm}).

In this section we will study the integrable structure of two-dimensional
type 0A string theory at the radius $R=\frac{l_s}{\sqrt{2}}$ (\ref{topp}).
Recently, the Toda integrable 
structure in the matrix model
formulation of type $0$A strings was proved in \cite{Yin:2003iv}.
We will study the matrix model of the system.

First,  
we will follow \cite{Dijkgraaf:1992hk} and derive Ward identities for the deformed
matrix model giving recursion relations for the tachyon correlators of the $0$A string.
We will then construct a Kontsevich-like matrix model 
representation of the generating functional of the tachyon correlators.
Both results will have an interpretation in terms of the topological 
description, which will be discussed in the next section.
For a review of matrix models and their integrable structure,
see for instance \cite{Dijkgraaf:1991qh,Mukhi:2003sz}.
The integrable structure of the deformed matrix model is investigated in
detail in \cite{Demeterfi:1993sj,Avan:hv,Nakatsu:1995eg}.

\subsection{Type $0$A and the Deformed Matrix Model}

% now with alpha^{\prime}=2 conventions...

As noted in \cite{Kapustin:2003hi,DeWolfe:2003qf}, the 
type $0$A matrix model \cite{Douglas:2003up} can be recast as the deformed 
matrix model \cite{Jevicki:1993zg}.

The  $U(N+q) \times U(N)$ type 0A matrix model,
which is the theory on the worldvolume
of $N+q$ $D0$-branes and $N$ $\bar{D0}$-branes of type 0A string theory,
can be described by $N \to \infty$ free fermions moving on the 
half line $\lambda \ge 0$,  whose dynamics is governed
by the Hamiltonian
\begin{equation}
\label{Hamilton1}
H=-\frac{1}{2}\frac{d^2}{d \lambda^2} 
-\frac{1}{2 \alpha^2}\lambda^2+\frac{q^2-\frac{1}{4}}{2\lambda^2} \ .
\end{equation}

The eigenvalues $\lambda$
of the matrix are  identified with the open string tachyon field
on the unstable D-branes \cite{McGreevy:2003kb}, which in type 0A is the tachyon
on the unstable $D0-\bar{D}0$ pair. 
In the original formulation of \cite{Douglas:2003up} the $0$A string is described
by free fermions moving in $2+2q$ dimensions in an inverted harmonic oscillator
potential. The curvature of the potential is identified with the mass
of the open string tachyon, which reads
\begin{equation}
\label{tachyon1}
m^2_T=-\frac{1}{2\alpha^{\prime}} \ .
\end{equation}
Thus, in (\ref{Hamilton1}) we have the relation between the
parameter $\alpha$ and the string length
according to
\begin{equation}
\alpha=\sqrt{2\alpha^{\prime}} \ .
\end{equation}
In the conventions we employ throughout the paper we set $\alpha^{\prime}=2$.

The reflection coefficient $R_p$ of the deformed matrix model
has been derived in \cite{Demeterfi:1993cm}. In our conventions
it is, up to a $p$ independent phase, given by
\begin{equation}
\label{ReflCoeff1}
R_p=\left(\frac{4}{q^2+4\mu^2-\frac{1}{4}}\right)^p
\frac{\Gamma\left(\frac{1}{2}+\frac{q}{2}+p-i\mu \right)}
{\Gamma\left(\frac{1}{2}+\frac{q}{2}- p+ i\mu  \right)} \ ,
\end{equation}
which is valid for $q \ge 0$ and $p > 0$.
The scattering amplitudes in the deformed matrix model
can be obtained by applying the rules of
\cite{Moore:1991zv}. For instance, 
the two-point function for the collective field
is given by
\begin{equation}
\label{Correlator1}
\mathcal{A}_2(p,-p)=\int_0^p dx R_{p-x}R^*_x \ .
\end{equation}
The collective field is related to the string theory
tachyon via leg pole factors (see appendix A).
%We will rescale the tachyon vertex operators by
%the cosmological constant in order to organize the 
%perturbation theory for the correlators in a convenient way.

\subsection{Ward Identities}

Consider the deformed matrix model compactified on a circle with radius
$\beta$.
This is equivalent to the  free fermions description at temperature
$T=\frac{1}{2\pi \beta}$. The allowed fermion momenta along the
time direction are $p_m=\frac{1}{\beta} \left(m+\frac{1}{2}
\right)$.
The discrete momenta for the bosonic collective field are $p_n=n/\beta$.

As in \cite{Dijkgraaf:1992hk} we will introduce two bosonic fields
$\partial \phi^{in/out}(z)=\sum_k \frac{\alpha^{in/out}}{z^{k+1}}$ such
that
\begin{equation}
\label{CorRel1}
\left<\prod_i^N V_{n_i/\beta} \prod_i^{N^{\prime}} V_{-n_i/\beta} \right>=
-\frac{(i \mu)^{N+N^{\prime}}}{\beta} 
\left<\prod_i^N \alpha_{n_i}^{out} \prod_i^{N^{\prime}} \alpha_{-n_i}^{in}
\right> \ .
\end{equation}
The $in/out$ fields are related by the $S$ operator 
\cite{Dijkgraaf:1992hk}.

We introduce the generating functional for the connected tachyon correlators 
\begin{eqnarray}
\mu^2 \mathcal{F} & = & 
\left< e^{\sum_{n>0} t_n V_{n/\beta}} 
e^{\sum_{n>0} \bar{t}_n V_{-n/\beta}} \right>_c  \nonumber \\
& & =-\frac{1}{\beta} \left<0 \right. \left. \right| 
e^{i \mu \sum_{n>0} t_n \alpha_n} S
e^{i \mu \sum_{n>0} \bar{t}_n \alpha_{-n}} \left| 0 \right>_c \ ,
\end{eqnarray}
now written in terms of the modes of a single free boson.
Introducing 
$Z=e^{\mu^2 \mathcal{F}}$, and the following
shorthand for the coherent states
\begin{equation}
\label{CohSt1}
\left|\bar{t} \right> \equiv e^{i \mu \sum_{n>0} \bar{t}_n \alpha_{-n}}
  \left| 0 \right> \ ,
\label{tcoh}
\end{equation}
we can write the derivative with respect to the coupling
$\bar{t}_n$ as
\begin{equation}
\frac{\partial \mathcal{F}}{\partial \bar{t}_n} = -\frac{i}{\mu}Z^{-1}
\left<t \right| \alpha_n S \left|\bar{t}\right> \ .
\end{equation}
Upon fermionization and applying the $S$ operator,
this can be
written as
\begin{equation}
\label{Formula1}
\frac{\partial \mathcal{F}}{\partial \bar{t}_n} = -\frac{i}{\mu}Z^{-1}
\oint \frac{dw}{w^n} \oint \frac{dz}{z}
\left[ \sum_{m \in Z}R^*_{p_m}R_{n/\beta-p_m} \left(\frac{w}{z}
  \right)^m \right]
\left<t \right|S \psi(z) \bar{\psi}(w) \left|\bar{t}\right> \ .
\end{equation}
The product of the reflection coefficients
simplifies considerably if we set $\beta=1$.
This is the radius (\ref{topp}).

We now have $p_m=\left(m+\frac{1}{2} \right)$, and  we get
\begin{eqnarray}
R^*_{p_m}R_{n-p_m} & = & \left(\frac{4}{q^2+4\mu^2
    -\frac{1}{4}} \right)^n \left(1+\frac{q}{2}+m+ i
      \mu-n \right)_n
\left(\frac{q}{2}-i \mu -m \right)_n \nonumber \\
 & =& \left(\frac{4}{q^2+4\mu^2
    -\frac{1}{4}} \right)^n(-1)^n \left[-1-\frac{q}{2}-m- i
      \mu +n\right]_n
  \left(\frac{q}{2}- i\mu -m\right)_n \nonumber \\
 &= & \left(\frac{4}{q^2+4\mu^2
    -\frac{1}{4}} \right)^n(-1)^n
  \left(-i\mu-\frac{q}{2}-m\right)_n
\left(-i\mu+\frac{q}{2}-m\right)_n \ .
\end{eqnarray}
We used the definitions
\begin{equation}
\label{GammaF1}
\Gamma(z+n)=(z)_n\Gamma(z) \ ,
\end{equation}
where $(z)_n$ is the Pochhammer symbol defined
by
\begin{equation}
\label{Poch1} 
(z)_n=z(z+1)...(z+n-1) \ .
\end{equation}
Also, 
\begin{equation}
[z]_n=z(z-1)...(z-n+1) \ ,
\end{equation}
and we have the identity
\begin{equation}
\label{ID1}
(z)_n=(-1)^n[-z]_n \ .
\end{equation}

In the sum of (\ref{Formula1}) we can replace the Pochhammer
symbols by a differential operator and after partial integration we
can transform
\begin{equation}
 \left(-i\mu-\frac{q}{2}-m\right)_n \to
\left(-i\mu-\frac{q}{2}-z \partial_z\right)_n \ ,
\end{equation}
where the derivative acts on $\psi(z)$ in (\ref{Formula1}).
Using the identity
\begin{equation}
\left(a-z\partial_z\right)_n= (-1)^nz^n z^a \left(\partial_z\right)^n
z^{-a} \ ,
\end{equation}
we find 
\begin{eqnarray}
\label{DOperator1}
& & \left(-i\mu-\frac{q}{2}-z \partial_z\right)_n
\left(-i\mu+\frac{q}{2}-z \partial_z\right)_n
\nonumber \\
& & = z^{n-\hat{q}}z^{-i\mu}\left( \partial_z \right)^n
z^{n+q}\left( \partial_z \right)^n z^{i \mu -\hat{q}} \
,
\end{eqnarray}
where for notational convenience we have introduced the hatted
quantity $\hat{q} \equiv \frac{q}{2}$.
As in \cite{Dijkgraaf:1992hk}, we bosonize $\psi(z)$ into a
field $\phi(z)$ and shift the zero mode 
\begin{equation}
\phi(z)=\tilde{\phi}(z) +\left(\hat{q}-i\mu \right) \log{z} \ ,
\end{equation}
to simplify the action of the operator (\ref{DOperator1}).
Taking the normal ordered product of the two exponentials,
expanding around $z=w$, and taking into account that
$\frac{1}{z}\sum_{m \in Z} \left(\frac{w}{z}\right)^m$ acts like a 
delta function we obtain
\begin{eqnarray}
\label{Ws1}
\frac{\partial \mathcal{F}}{\partial \bar{t}_n} & = &  
(-1)^{n+1} \frac{i}{\mu} Z^{-1}
\left(\frac{4}{q^2+4\mu^2-\frac{1}{4}} \right)^n 
\sum_{l=0}^n
\frac{1}{2n+1-l} {n \choose l}\left[ n+ q \right]_l \times 
\nonumber \\
& & \times \oint dw w^{n-l}:e^{-i \mu \varphi(w)} \left(\partial_w
  \right)^{2n+1-l} e^{i\mu \varphi(w)}: Z \ ,
\end{eqnarray} 
where the rescaled operator $\varphi$ has a representation
on the coherent states as
\begin{equation}
\label{CSR1}
\partial \varphi(w)=\frac{1}{w} \left(1+ i\frac{\hat{q}}{\mu}
\right)+ \sum_{k\ge 1} k t_k w^{k-1} - \frac{1}{\mu^2} \sum_{k \ge 1}
\partial_{t_k}w^{-k-1} \ .
\end{equation}

The standard generators of the $W_{1+\infty}$ algebra are 
written in terms of the operators of the form
\begin{equation}
P^{(n)}=:e^{-\tilde{\phi}(z)} \left(\partial_z \right)^n e^{\tilde{\phi}(z)}: \ ,
\end{equation}
and derivatives thereof as
\cite{Pope:1989sr}\cite{Fukuma:1990yk}
\begin{equation}
W^{(n)}(z)= \sum_{l=0}^{n-1}\frac{(-1)^l}{(n-l)
  l!}\frac{\left(\left[n-1\right]_l\right)^2}{\left[2n-2\right]_l}
  \partial_z^lP^{(n-l)}(z) \ .
\end{equation}

If we introduce the operators
\begin{equation}
W^{(k)}(z, q)=
\sum_{l=0}^{k-1}\frac{(-1)^l}{(k-l)l!}\left[\frac{k-1}{2}+q \right]_l
\partial_z^l P^{(k-l)}(z) \ ,
\end{equation}
we can write
\begin{eqnarray}
:W^{(2n+1)}_{-n}(q):Z & = & \oint dz z^n
:W^{(2n+1)}(z,q):Z \nonumber \\
& = & \sum_{l=0}^{2n}\frac{(-1)^l}{(2n+1-l) l!} \left[n+q
\right]_l \oint dz z^n \partial^l_z :P^{(2n+1-l)}(z):Z \nonumber \\
& = & \sum_{l=0}^{2n}\frac{(-1)^l}{(2n+1-l) l!} \left[n+q
\right]_l \oint dz (-1)^l \left(\partial_z^l z^n\right) :P^{(2n+1-l)}(z):Z \nonumber \\
& = & \sum_{l=0}^{n}\frac{\left[n+q
\right]_l}{(2n+1-l)}{n \choose l}  \oint dz  z^{n-l}
:P^{(2n+1-l)}(z):Z \ .
\end{eqnarray}

It would be interesting to check the algebra of 
 $W^{(n)}(q)$ and its relation to the standard form of the  $W_{\infty}$
algebra.
Note that for $q=0$, 
we expect only 1/2 of the $W_{\infty}$
  algebra of the $c=1$ model  \cite{Demeterfi:1993sj}.
Indeed in this case, we only 
  have odd spin generators in our algebra.

\subsection{A Kontsevich-like Matrix Model}

In this section we will derive a $0+0$ dimensional Kontsevich-like
 {\it two} matrix model, which describes type 0A strings at the
radius  $R=\sqrt{\frac{\alpha^{\prime}}{2}}=1$ (\ref{topp}).

\subsubsection{The Strategy}

Our derivation will follow the approach taken in \cite{Dijkgraaf:1992hk} and  
\cite{Imbimbo:1995yv}
for the $c=1$ bosonic string at the self-dual radius.
The generating functional for
the tachyon correlators was expressed as the expectation value
of the operator  $S$,
\begin{equation}
\label{GenF1}
Z(t, \bar{t})=\left< t \right. \left|S \right| \bar{t} \left. \right> \ ,
\end{equation}
where the state $\left| t \right. \left. \right>$ is a coherent state
of a holomorphic boson with the expansion $\partial \varphi(z)
=\sum_n \alpha_n z^{-n-1}$
(\ref{tcoh}).

As in the bosonic case, $\varphi$ may be thought of as
the NS-NS tachyon at spatial infinity with periodic Euclidean time.
We make contact with the two dimensional 0A string theory by writing 
\begin{equation}
\label{GenF2}
Z(t, \bar{t})=\left<e^{\sum_{n \ge 1}t_n V_{n/R}+ 
\sum_{n \ge 1}\bar{t}_n V_{-n/R}} \right> \ ,
\end{equation}
where $V_{n/R}$ are the tachyon vertex operators with (Euclidean)
time compactified at radius $R$.

The evaluation of (\ref{GenF1}) proceeds by
expressing the two-dimensional boson $\varphi$ using
fermions
\beq
\partial \varphi(z)=
:\bar{\psi}(z) \psi(z): \ .
\eeq
In this language the
action of the operator $S$ is encoded in the
reflection coefficients $R_{p_n}$ 
\begin{equation}
\label{Saction1}
S\psi_{-n-1/2}S^{-1}=R_{p_n}\psi_{-n-1/2} \ ,
\end{equation}
where $p_n=\frac{1}{R}\left(n+\frac{1}{2}\right)$ is the
fermionic momentum along the compact direction.
Here, the reflection coefficients $R_{p_n}$
are given by (\ref{ReflCoeff1}).
It is convenient to represent
the fermionic Fock space in terms of semi-infinite forms.
The vacuum $\left| 0 \right. \left. \right>$ in this
case is written as
\begin{equation}
\label{FVacuum1}
\left| 0 \right. \left. \right>=z^0 \wedge z^1 \wedge z^2 \wedge ... \ .
\end{equation}
The operators $\psi_{n+1/2}, \bar{\psi}_{n+1/2}$ 
annihilate the vacuum for positive $n$, and
are represented by
\begin{equation}
\label{Repr1}
\psi_{n+1/2}=z^n, \, \;\; \bar{\psi}_{-n-1/2}=\frac{\partial}{\partial z^n} \ .
\end{equation}
Hence, $S$ acts as
\begin{equation}
\label{Saction2}
S : z^n \to R_{-p_n} z^n \ .
\end{equation}

The operator $U(\bar{t})$ that creates
the coherent state (\ref{tcoh}) from the vacuum, acts on the semi-infinite
forms as
\begin{equation}
U(\bar{t}): z^n \to e^{i \mu \sum\bar{t}_k z^{-k}}z^n=\sum_{k=0}^{\infty}
P_k(i \mu \bar{t})z^{n-k} \ ,
\label{shur}
\end{equation}
where $P_k(\bar{t})$ are the Schur polynomials. 
The combined action of $S$ and $U(\bar{t})$ reads
\begin{equation}
\label{CA1}
S \circ U(\bar{t}): z^n \to w^{(n)}(z,\bar{t})
=\sum_{k=0}^{\infty}P_k(i \mu \bar{t})R_{-p_{n-k}}z^{n-k} \ ,
\end{equation}
such that
\begin{equation}
\label{St1}
S \left| \bar{t} \right. \left. \right> = w^{(0)}(z,\bar{t}) \wedge 
w^{(1)}(z,\bar{t})\wedge ... \ .
\end{equation}

In order to  express the bra coherent state $\left< t \right. \left. \right|$, 
we need the Miwa transformation, which relates
the sources $t_n$ to a constant matrix $A$ by
\begin{equation}
\label{Miwa1}
i \mu t_n=-\frac{1}{n} \mathbf{Tr} A^{-n} \ .
\end{equation}
Denoting the eigenvalues of the $N \times N$ matrix $A$
by $a_i, \, i=1,..,N$ we can write
\begin{equation}
\label{CST2}
\left< t \right. \left. \right|=\left< 0 \right. \left. \right|
\prod_{i=1}^{N}e^{-\sum_{n>0}\frac{1}{n}\frac{\alpha_n}{a_i^n}}=
 \left< N \right. \left. \right|\frac{\prod_{i=1}^N\psi(a_i)}{\Delta(a)} \ ,
\end{equation}
where $\Delta(a)$ denotes the Vandermonde determinant and 
\begin{equation}
\label{NFermi1}
\left. \right| N \left> \right. = z^N \wedge z^{N+1} \wedge z^{N+2} \wedge ... 
\end{equation}
denotes the $N$-fermion state.
We now
put together (\ref{St1}) and (\ref{CST2}) and get
\begin{equation}
\label{Res1}
Z(t,\bar{t})=\left<t \right. \left|S \right| \bar{t} \left> \right.=
\frac{\mathrm{det} w^{(j-1)}(a_i)}{\Delta(a)} \ .
\end{equation}

Thus, the derivation is exactly the same as
in the bosonic case, 
where the distinctive features of the system enter
in the reflection coefficients $R_{p_n}$.

\subsubsection{The Derivation}

Let us compute first when the RR
background flux is set to zero $q=0$.
The exact reflection coefficients are given by 
\begin{equation}
\label{ReflCoeff0m}
R_p=K^{p-i\mu}e^{-i2\mu}\frac{\Gamma\left(\frac{1}{2}+p-i \mu \right)}
{\Gamma\left(\frac{1}{2}-p+i \mu \right)} \ ,
\end{equation}
with
\begin{equation}
\label{K2}
K=\frac{4}{4\mu^2-\frac{1}{4}} \ .
\end{equation}
We will consider 
only the perturbative properties of the reflection coefficients.
Using Gamma function identities, we can rewrite the reflection coefficient
$R_p$ as
\begin{equation}
\label{ReflCoeff0m2}
R_p=\frac{K^p}{2\pi}e^{i \pi p}e^{\pi \mu}\left[\Gamma
\left(\frac{1}{2}+p- i \mu \right)\right]^2 \ ,
\end{equation}
up to terms $O\left(e^{-\mu} \right)$ and a $p$ independent phase.

To calculate the one particle states $w^{(n)}(z,\bar{t})$
we plug (\ref{ReflCoeff0m2}) into (\ref{CA1})
and use the integral representation of the Gamma function.
This yields
\begin{eqnarray}
\label{CA2}
& & w^{(n)}(z,\bar{t})=-\frac{i}{2\pi}e^{-i2\mu}e^{-i\pi n}e^{\pi \mu }K^{-i\mu-n-1/2}z^n \\
& & \times \int_0^{\infty}dm_1 dm_2\left(m_1 m_2 \right)^{-n-i\mu-1}
e^{-m_1-m_2}e^{i \mu \sum_{k>0}\bar{t}_k\left(-K \right)^k
\left(\frac{m_1m_2}{z}\right)^k} \nonumber \\
& & =-\frac{i}{2\pi}e^{-i2\mu}K^{-1/2}z^{-i\mu}
\int_0^{\infty}dm_1 dm_2\left(m_1 m_2 \right)^{-n-i\mu-1}
e^{-m_2}e^{\frac{z}{K}m_1}e^{i \mu \sum_{k>0}\bar{t}_k\left(m_1m_2\right)^k} \ . \nonumber
\end{eqnarray}
Here we used the relation (\ref{shur}), that involves the Schur polynomial.
Introduce now two new integration variables by $m=m_1m_2,\, s=m_1$.
We have
\begin{eqnarray}
\label{CA3}
w^{(n)}(z,\bar{t})=-\frac{i}{2\pi}e^{-i2\mu}K^{-1/2}z^{-i\mu}
\int_0^{\infty}ds s^{-1} e^{\frac{z}{K}s} \int_0^{\infty}dm m^{-n-i\mu-1}
e^{-m/s}e^{i \mu \sum_{k>0}\bar{t}_k m^k} \ .
\end{eqnarray}
Evaluating the expression (\ref{Res1}), we get
\begin{eqnarray}
\label{DetA1}
\frac{\mathrm{det}\, w^{(j-1)}(a_i,\bar{t})}{\Delta(a)} & = &
c(\mu)^N \prod_i a_i^{-i \mu} \prod_i\left( 
\int_0^{\infty}ds_i s_i^{-1} e^{\frac{a_i}{K}s_i} \int_0^{\infty} dm_i m_i^{-i\mu-1}
e^{-m_i s_i^{-1}}e^{i \mu \sum_{k>0}\bar{t}_k m_i^k} \right)
\times \nonumber \\
& &  \times \frac{\mathrm{det} \, \left(m_i^{-1}\right)^{j-1}}{\Delta(a)} \ ,
\end{eqnarray}
where $c(\mu)=-\frac{i}{2\pi}e^{-i2\mu} K^{-1/2}$.
Note that
\begin{equation}
\det \left(m_i^{-1}\right)^{j-1}=\Delta(m^{-1})
={\Delta(m)\over\prod_i m_i^{N-1}} \, .
\end{equation}
To rewrite the integral expression in terms of a matrix integral, we use
the Harish-Chandra/Itzykson-Zuber formula 
\cite{Harish-Chandra,Itzykson:1979fi}
\begin{equation}
\label{HC1}
\int d\Omega_{U(N)}e^{\mathrm{Tr} \left(\Omega^{\dagger}x\Omega y\right)}
=\prod_{k=1}^{N-1}k!\frac{\mathrm{det} \,e^{x_iy_j}}{\Delta(x) \Delta(y)} \ ,
\end{equation}
where $x,y$ are diagonal matrices.
{}From this formula, one can verify that for any symmetric function of
$g(x_i)$
\begin{equation}
\int [dX] e^{\tr (XY) }g(X)
\sim \int \prod_{i=1}^Ndx_i\, {\Delta(x)\over\Delta(y)}\,e^{\sum_i x_iy_i}
g(x) \, ,
\end{equation}
with
\begin{equation}
X=\Omega_{U(N)}x\,\Omega_{U(N)}^{\dagger},~~~
Y=UyU^{\dagger} \,~~ {\rm for~any}~U\in U(N) \, .
\end{equation}
It then follows that
\begin{eqnarray}
\label{MI1}
Z(t,\bar{t}) &\!\sim\!& c(\mu)^N \left(\mathrm{det}\, A\right)^{-i\mu}
\int \prod_{i=1}^N
\left({ds_i\over s_i^N}\,e^{{a_i\over K}s_i}\right)
{\Delta(s)\over\Delta(a)}\,
\int [dM]\left(\det M\right)^{-{i\mu}-N}
e^{i\mu\sum_k\bar{t}_k\tr M^k-\tr (MS^{-1})}
%\,e^{-\tr (MS^{-1})}
\nn
&\!\sim\!&
c(\mu)^N K^{-\frac{N\left(N-1\right)}{2}}\left(\det A\right)^{-i\mu}
\int \left[dM_1\right]  
e^{\frac{1}{K}\mathrm{Tr} \left(M_1A\right)-N \tr\log M_1}\nn
%e^{-N \mathrm{Tr\, log}\,M_1} 
&&
\qquad\qquad\qquad
\int \left[dM_2\right]
e^{-\mathrm{Tr} \left(M_1^{-1}M_2\right)}
e^{-\left(i \mu+N\right) \mathrm{Tr\, log}\,M_2}
e^{i \mu\sum_{k>0}\bar{t}_k \mathrm{Tr}\, M_2^k} \ ,
\end{eqnarray}
where by $\sim$ we mean up to a constant factor.
This is a  Kontsevich-Penner like
 {\it two} matrix model over positive definite matrices, which describes type 0A strings at the
radius  $R=1$. The convergence properties of this integral are similar to those of the Imbimbo-Mukhi
matrix integral \cite{Imbimbo:1995yv}, which derive directly from the properties of the
$\Gamma$-functions.

Consider next the case of
non vanishing RR background flux $q\neq 0$.
Instead of (\ref{ReflCoeff0m}), we use now the reflection
coefficients for finite RR background flux (\ref{ReflCoeff1}),
which we can rewrite in perturbation theory as
\begin{equation}
\label{ReflCoeff3}
R_p=\frac{\tilde{K}^{p-i\mu}}{2\pi}e^{i \pi p-i2\mu}e^{\pi \mu}\Gamma
\left(\frac{1}{2}+\frac{q}{2}+p-i \mu \right)
\Gamma \left(\frac{1}{2}-\frac{q}{2}+p-i \mu \right) \ .
\end{equation}
Following the same steps as above,
we find that we can write the generating functional of
the momentum mode tachyon correlators as
\begin{eqnarray}
\label{MI2}
Z(t,\bar{t}) & \sim & \tilde{c}(\mu)^N \tilde{K}^{-\frac{N(N-1)}{2}} 
\left(\mathrm{det}\, A\right)^{\frac{q}{2}-i\mu}
\int \left[dM_1\right]  e^{\frac{1}{\tilde{K}}\mathrm{Tr} \left(M_1A\right)}
e^{-(N-q) \mathrm{Tr\, log}\,M_1}\int \left[dM_2\right]
e^{-\mathrm{Tr} \left(M_1^{-1}M_2\right)} \nonumber \\
& & e^{-\left(q/2+i \mu+N\right) \mathrm{Tr\, log}\,M_2}
e^{i \mu\sum_{k>0}\bar{t}_k \mathrm{Tr}\, M_2^k} \ ,
\end{eqnarray}
with $\tilde{c}(\mu)=\frac{i}{2\pi}\left(-1\right)^{q/2+1}e^{-i2\mu}\tilde{K}^{q/2-1/2}$, and
$\tilde{K}=\frac{4}{q^2+4\mu^2-1/4}$.
It is obvious that (\ref{MI2}) reproduces (\ref{MI1}) in the case $q=0$.
It would be interesting to generalize the matrix model to other radii by using a normal matrix
model as was done for the $c=1$ string in \cite{Alexandrov:2003qk}.

\subsubsection{The Partition Function}

We now set $\bar{t}$ to zero in order to get the unperturbed partition
function.  Consider the case $q=0$.
We can write $Z$ as a product of two independent
matrix integrals
\begin{eqnarray}
\label{PFA1}
Z(t=\bar{t}=0) & \sim & {\mathcal A}(\mu) {\mathcal N}^{-2}
\int \left[dM_1\right] e^{\frac{1}{K}\mathrm{Tr} \left(M_1\right)
{-\left(i \mu+N\right) \mathrm{Tr\, log}\,M_1}} \times \nonumber \\ 
& & \times \int \left[dM_2\right]e^{-\mathrm{Tr} \left(M_2\right)
{-\left(i \mu+N\right) \mathrm{Tr\, log}\,M_2}} \ ,
\end{eqnarray}
with ${\mathcal A}(\mu) \sim K^{i\mu N+\frac{N^2}{2}}, 
{\mathcal N} = \int dMe^{-N\tr \frac{1}{2}M^2}$.
After rescaling and shifting the integration variables,
$Z$ is written as a square of the Penner integral
\begin{equation}
\label{PI1}
Z \sim {\mathcal N}^{-2}\left(\int dMe^{-Nt\sum_{k>2}\mathrm{Tr}\frac{M^k}{k}}\right)^2 \ ,
\end{equation}
where $t=-\left(1+\frac{i \mu}{N}\right)$.
In the limit of large $N$, we reproduce the partition function of the $0$A
theory compactified at $R=1$ in units of $\alpha^{\prime}=2$ 
\cite{Douglas:2003up}.

\section{Topological B-Model}
\label{Section6}

The  topological B-model is a twisted $N=2$ supersymmetric $\sigma$-model.
Its marginal deformations are complex 
deformations of the target space \cite{Witten:1991zz}. Thus, 
the topological B-model studies the complex deformations of the target
space, which is known as the Kodaira Spencer theory \cite{Bershadsky:1993cx}.
The basic field of the B-model $A \in H^{(0,1)}(T^{1,0}M)$
is called the Kodaira-Spencer Field.

Recently, a simplified framework for studying the topological 
B-model on non-compact Calabi-Yau manifolds $M$ of the form
\beq
uw - H(x,y) = 0 \ ,
\label{uw}
\eeq
has been suggested in 
\cite{Aganagic:2003qj}.
In this section we would like to 
 consider some aspects of $\hat{c}=1$ strings 
via a topological description in this framework.

The ground ring of 
superaffine 0B at the self-dual radius takes the form (\ref{uw}) with
$H(x,y) = xy -\mu$. This case has been studied in detail in
\cite{Aganagic:2003qj} in the context of the $c=1$ bosonic string at the self-dual radius.
On the other hand the  ground ring of 
superaffine 0A at the self-dual radius does not take the form (\ref{uw}), and we cannot study
it in this simplified setting.
The (singular) ground rings of type 0A (0B) strings at the radii (\ref{topp}) take the form (\ref{uw})
and we will consider this case.
In addition, we will consider for comparison the $Z_n$ orbifolds of the bosonic $c=1$ string.

The
observables correspond to variations
of the complex structure at infinity, and will be identified
with the momentum modes of the tachyon field. 
In fact, one considers only deformations of 
$H(x,y)$, keeping the dependence on $u,w$ fixed.
One can think of the deformations of the Riemann surface $H(x,y)=0$
as corresponding to the deformations of the Fermi surface
in the matrix model by the tachyon perturbations.
In this case,  
the problem reduces
to a one complex-dimensional one
\beq
\int_{3-cycle} \Omega \rightarrow \int_{1-cycle} y(x) dx \ .
\label{red}
\eeq
The complex structure deformations of 
$H(x,y)$ are encoded in the one-form
\beq
\lambda =y\,dx \ .
\label{lam}
\eeq
In each (asymptotic) patch the variations of the complex structure
are described
by the modes of a chiral boson $\phi(x)$, defined by
\beq
y(x)=\partial \phi(x)= y_{cl}(x) + t_0 x^{-1} + \sum_{n>0} n t_n
x^{n-1} +
\sum_{n>0} {\cal F}_n x^{-n-1} \ .
\eeq
 $\phi(x)$ is the Kodaira-Spencer field, and 
 $y_i^{\rm cl}(x_i)$ is the solution of $H(x,y)=0$.
We will associate the commutator of $x$ and $y$ with the string equation.

One defines a 
fermion field
\begin{equation}
\psi(x)=e^{\frac{{\phi}(x_i)}{g_s}} \ .
\label{fermion}
\end{equation}
As outlined in \cite{Aganagic:2003qj} non-compact branes fibered over the Riemann surface,
whose backreaction changes the complex structure, can be described by these essentially
free fermions, which, however, do not transform geometrically from patch to patch
\footnote{A similar geometrical interpretation of D-branes
in the non-critical string has been recently given in
\cite{Seiberg:2003nm} (see also \cite{Alexandrov:2004ks}\cite{Kazakov:2004du})}.
Rather it was proposed in \cite{Aganagic:2003qj} that they transform according to 
\begin{equation}
\psit(\xt)=\int dx \,e^{-S(x,\xt)/g_s}\,\psi(x) \ ,
\label{cano:fermion}
\end{equation}
with $S(x,\xt)$ a canonical transformation relating the $x/\tilde{x}$ patches.
For instance, for the $c=1$ string at the self-dual radius, one has the familiar Fourier
transformation with \cite{Aganagic:2003qj}
\begin{equation}
S(x,\xt)=x\xt \ .
\end{equation}

\subsection{Partition Function and Geometry of Type $0$A Strings}

In this section we will study the Calabi-Yau B-model geometry 
corresponding to the type 0A strings at the radius (\ref{topp}).
We will see how the B-model reproduces precisely the partition function 
of the type 0A strings, even when the RR flux $q$ is different than zero.
We will start by considering for comparison,
  the  $Z_n$ orbifolds of the $c=1$ bosonic string 
\cite{Gopakumar:1998vy}, and consider next the  type 0A strings.
Previous work on the connection between the deformed matrix model
and topological field theory appeared in \cite{Danielsson:1994ac}.

\subsubsection{$Z_n$ Orbifolds of the $c=1$ Bosonic String}

In \cite{Gopakumar:1998vy} it was argued that the B-model geometry
corresponding to the $c=1$ bosonic string at a radius $R=R_{sd}/n$, with $n$ integer and
$R_{sd}$ the self dual radius should arise from an orbifold of the
conifold in the following way. The $Z_n$ orbifold group acts on the 
$c=1$ ground ring
elements $x_{ij}$ (\ref{con}) as 
\begin{equation}
\left\{x_{11},x_{22},x_{12},x_{21} \right\} \to 
\left\{\omega x_{11},\omega^{-1}x_{22},x_{12}, x_{21} \right\} \ ,
\end{equation}
where $\omega^n=1$ and $x_{11},x_{22}$ are the momentum modes.

The deformed $c=1$ ring relation (\ref{cond})
can be written as 
\begin{equation}
x_{11}x_{22}=t \; , \;t=x_{12}x_{21}+\mu \ .
\end{equation}
Modding out by $Z_n$ leads to an $A_{n-1}$ singularity, which in terms of  
the invariant coordinates $w=x_{11}^n,z=x_{22}^n$ is
\begin{equation}
w z=t^n  \; , \;t= x_{12}x_{21}+\mu \ .
\end{equation}
If we deform the $A_{n-1}$ singularity we get a 
smooth geometry
\begin{equation}
\label{GeomO1}
wz=\prod_i \left(x_{12}x_{21}  - \hat{\mu}_i \right) \ ,
\end{equation}
with $\hat{\mu}_i=\mu_i-\mu$. 

We have $n$ inequivalent 
3-cycles in this geometry, whose sizes are determined
by the parameters $\hat{\mu}_i$. It was suggested in
\cite{Gopakumar:1998vy}, that the partition function
of the topological B-model on this geometry, describing the 
nearly massless hypermultiplets arising from D-branes
wrapping the $n$ shrinking cycles, should map to the
partition function of the $Z_n$ orbifold of the
$c=1$ bosonic string for appropriate $\hat{\mu}_i$.
Each of the cycles gives a contribution
equal to the conifold one, such that the partition
function $\cF(\mu)$ of the $Z_n$ quotient  
can be written as
\begin{equation}
\label{Oc1}
\cF(\mu)=\sum_i {\mathcal F}_{c=1} (\hat{\mu}_i) \ ,
\end{equation}
where ${\mathcal F}_{c=1} (\hat{\mu}_i)$ is the partition function
of the $c=1$ string at the self dual radius.
 
Indeed, we can write the free energy of the $c=1$ string
at radius $R$ as (see e.g. \cite{Kostov:2002tk})
\begin{equation}
\label{FER1}
\cF(\mu,R)=\int_{-\infty}^{\infty}dE \rho(E) \log{\left[1+e^{-\beta\left(\mu+E\right)}\right]} \ ,
\end{equation}
where $\beta=2\pi R$. If we use the fact that the density of states $\rho(E)$ is related to the
reflection coefficient $R(E)$ by
\begin{equation}
\rho(E) \sim \frac{d \phi(E)}{dE} \; , \; \phi(E) \sim \log{R(E)} \ ,
\end{equation}
we can integrate by parts and pick up the residues in the upper half plane
in (\ref{FER1}) to obtain
\begin{equation}
\cF(\mu,R)=-i \sum_{r \in Z_++1/2}\phi(ir/R-\mu) \ . 
\end{equation}
Plugging the reflection coefficients $R(E) \sim \Gamma(iE +1/2)$ for the
$c=1$ string we arrive at
\begin{equation}
\cF(\mu,R)=\sum_{r,s \in Z_++1/2}\log(s+r/R-i\mu) \ .
\end{equation}    
As shown in \cite{Gopakumar:1998vy} for $R=1/n$ (note that our conventions
are $\alpha^{\prime}=1$), this can be recast in the form
\begin{equation}
\cF(\mu,R)=\sum^{\frac{n-1}{2}}_{k=-\frac{n-1}{2}} 
{\mathcal F}_{c=1} (\hat{\mu}_k) \ ,
\end{equation}
with $\hat{\mu}_k=\frac{\mu+ik}{n}$, thus fixing the deformation  parameters of 
the geometry (\ref{GeomO1}).

\subsubsection{The Type $0$A String}

We now apply the above arguments to the case
of the $0$A string. For our current purposes it will
be sufficient to know that the perturbative reflection coefficients
of the $0$A string (\ref{ReflCoeff3}) can be written as
\begin{equation}
R(E) \sim \Gamma^2\left(\frac{iE}{2}\alpha +\frac{q+1}{2} \right) \ ,
\ \alpha= \sqrt{2\alpha^{\prime}} \ .
\end{equation}
In our conventions of $\alpha^{\prime}=2$ for the fermionic string, this gives
for the partition function
\begin{equation}
\cF(\mu,R=1)=2 {\mathcal Re} \sum_{r,s \in Z_++1/2}
\log{\left(r+s-i \mu +\frac{q}{2} \right)} \ ,
\end{equation}
which upon comparison to the conifold partition function leads to the perturbative
identification
\begin{equation}
\label{C1vs0A}
\cF(\mu,R=1)={\mathcal F}_{c=1}\left(\mu+\frac{iq}{2}\right)
+{\mathcal F}_{c=1}\left(\mu-\frac{iq}{2}\right) \ .
\end{equation}
The corresponding geometry is (\ref{Geomy11}).

Note that in (\ref{C1vs0A}) we have chosen a linear combination that respects 
the discrete symmetry $q \to -q$ of the Hamiltonian (\ref{Hamilton1}), 
naturally leading to a geometry (\ref{Geomy11}) with the same symmetry.
Furthermore, setting $q=1/2$, which, of course, is unphysical in the $0$A string,
 and working with the conventions 
of $\alpha^{\prime} =1/2$ we can recover
the perturbative partition function of the $Z_2$ orbifold of the $c=1$ string (with $\alpha^{\prime}=1$).
This is expected since for $q=1/2$ the deformation parameter  
of the deformed matrix model (\ref{Hamilton1}) vanishes,  and the different curvature
of the inverted harmonic oscillator in the case of the fermionic string
translates into the appropriate $\alpha^{\prime}$ conventions.

The geometry (\ref{Geomy11}) is consistent with the 
undeformed ground ring of the 0A string (\ref{GR0AB1}), and gives a prediction 
for the deformation of the ground ring induced by the cosmological constant $\mu$ and 
the RR charge $q$.

In the geometry (\ref{Geomy11}) we have two independent parameters $\mu$
and $q$ 
describing deformations of the complex structure. In the neighbourhood
of a singularity in the complex structure moduli space the B-model
on (\ref{Geomy11}) is dominated by a universal behaviour determined
by the singularity. We suggest that different perturbative
expansions of the full non-perturbative solution of the $0$A string
describe the B-model on (\ref{Geomy11}) near different singularities in the
complex structure moduli space, where generally different 3-cycles
shrink to zero size. For instance, 
the perturbative expansion in $\mu \; (q)$ (for the 
asymptotic expansion in $q$ see e.g. \cite{Jevicki:1993zg}\cite{Demeterfi:1993sj} and 
\cite{Gukov:2003yp} for a recent discussion), which comprises the terms
that diverge in the limit $\mu \to 0 \; (q \to 0)$, should
describe the (perturbative) B-model near the points $\mu \to 0 \; 
(q \to 0)$ in the complex structure moduli space. 
In the following, we will discuss the point $\mu \to 0$ with $q=0$ held fixed.
We will be able to recover the integrable structure that describes
the $0$A tachyon scattering from the topological B-model.

\subsection{Reduction of the 3-Form}

As discussed above, 
the tachyons 
of the type $0$A string are associated  with the complex  
deformations of the  Riemann surface $H(x,y) = 0$.
A crucial ingredient of this correspondence is the reduction of the  
Calabi-Yau geometry to the Riemann surface (\ref{red}).
In this section
we will calculate the effective one-form on the Riemann surface (\ref{lam}). In 
the $0+1$-dimensional matrix model description, it is spanned by the momentum modes
of the ground ring.
We will derive the canonical transformation $S$, which relates
the two asymptotic patches of the Riemann surface.

We will first consider the B-model on the geometry (\ref{Geomy11}) for $q=0$.
For simplicity we will denote $u=a_{12}, v=a_{21}, x=b_{11}, y=b_{33}$.
Thus, 
\begin{equation}
\label{Geomy1}
\left(uv+\mu\right)^2=xy \ .
\end{equation}
The holomorphic 3-form $\Omega$
takes the form 
\beq
\Omega=\frac{dx}{x}\wedge du \wedge dv \ .
\eeq
We consider the integral of the three-form $\Omega$ over the
three-cycles of the form $S^1$ in the $u,v$ plane,
$u \to e^{i \theta} u, v \to e^{-i \theta} v$, fibered over a two real dimensional
disc in the $x,y$ plane bounded by a 1-cycle on the 
Riemann surface 
\begin{equation}
xy-\mu^2=0 \ .
\end{equation}
Over the Riemann surface the fiber develops a node at $u=v=0$, similar to the
case of $c=1$, but there is also a disconnected component that remains
regular.

Since the Riemann surface is a subset of the $x,y$ plane we replace
\begin{equation}
du \sim \frac{x dy}{2v \left(uv+\mu \right)}  \ ,
\end{equation}
and the 3-form may be written as
\begin{equation}
\Omega = \frac{1}{2} dx \wedge \frac{dy}{\sqrt{xy}}\wedge \frac{dv}{v} \ .
\end{equation}
Integrating over the circle in the fiber gives an effective 2-form in the $x,y$ plane,
\begin{equation}
\label{Eff2Form}
\int_{fiber}\Omega \to \frac{dx \wedge dy}{\sqrt{xy}} \ .
\end{equation}
The integral over the disk
in the $x,y$ plane can be rewritten by using Stokes theorem as an integral
of a one-form over the boundary 
\begin{equation}
\int \Omega \to \int_{\partial D}\frac{\sqrt{xy}}{y}dy \ ,
\end{equation}
such that we remain with an integral over a 1-cycle on the
Riemann surface $xy=\mu^2$,
\begin{equation}
\int \Omega \to \oint \mu\frac{dy}{y} \ .
\end{equation}

It is instructive to perform the reduction in a slightly different way.
We rewrite the geometry as
\begin{equation}
t=uv+\mu \, , t^2=xy  \ .
\end{equation}
In these coordinates, the three-cycle is given by
two circles in the $u,v$ and $x,y$ fibers, respectively,
fibered over an interval $t \in [0,\mu]$ in 
the $t$ plane. Reducing the 3-form 
\beq
\Omega=\frac{dx}{x} \wedge du \wedge dv = 
\frac{dx}{x} \wedge \frac{du}{u} \wedge dt \ , 
\eeq
to 
a 1-form in the $x,y$ fiber gives
\beq
\int \Omega \to \int \mu  \frac{dx}{x} \ ,
\eeq

Note that if we change $x \to y$ we get an additional minus sign.
This gives the one-form in the second asymptotic patch 
parametrized by $y$. 
In addition, the integral gives the deformation parameters
$\mu$ as expected.
The integration contour is a circle around the origin
in the complex $y$-plane. We might think of this contour as corresponding to the
compact one-cycle on the Riemann surface giving rise to the compact three-cycle
in the full Calabi-Yau \footnote{There is also a non-compact one cycle
corresponding to a different real section of the Riemann surface \cite{Alexandrov:2003qk}.}.

\subsection{Type 0A Fermionic Transformation}

As we have argued above the type $0$A string at radius $R=1$
corresponds to Kodaira-Spencer theory on the Riemann surface 
\begin{equation}
\label{RS22}
xy-\mu^2=0 \ ,
\end{equation}
which is part of the non-compact Calabi-Yau 
\begin{equation}
\left(uv+\mu\right)^2=xy \ .
\end{equation}
The Riemann surface has the topology of a sphere with two punctures.
The two punctures correspond to the asymptotic regions, whose 
complex deformations are mapped to the in- and outgoing $0$A tachyons,
respectively.

By pulling the contour along the sphere we can relate the action
of an operator in one asymptotic patch to the action in the
second asymptotic patch. In this way we can, for instance, write the
Ward identities (\ref{Ws1}) as fermionic bilinears \cite{Aganagic:2003qj} 
\begin{equation}
\label{Ws22}
\int_x \psi^*(x) x^n \psi(x) \left|V\right> =
-\int_y \psi^*(y) x^n(y) \psi(y) \left|V\right> \ .
\end{equation}
The state $\left|V\right>$ is in the Hilbert space ${\cal H}^{\otimes 2}$,
where ${\cal H}$ is the Hilbert space of a single free boson and
it is the quantum state of the Calabi-Yau in the Kodaira-Spencer theory. 

We now have to find the translation rule between the two patches.
From the previous reduction of the holomorphic 3-form we see that
$x,y$ are not canonically conjugate, in contrast to the cases discussed in
\cite{Aganagic:2003qj}. Instead, the 2-form (\ref{Eff2Form}) would rather
suggest a commutator $\left[x,y\right]=\sqrt{xy}$, which gives
rise to quantum ambiguities due to operator ordering. We will fix the ambiguities
by writing $x,y$ as operators on the Riemann surface, and requiring that 
they reproduce the equation of the Riemann surface (\ref{RS22}) in the classical 
limit. In other words we will go, say, to the $y$-patch, put a brane
at a fixed value $y_0$ and measure the corresponding $x$ value, which
in the classical limit is fixed by the equation of the Riemann surface
(\ref{RS22}). 

According to the proposal of \cite{Aganagic:2003qj}, the non-compact branes
fibered over the Riemann surface are described by fermions $\psi(z)$, which
can be bosonized as $\psi(z)=e^{-i\phi(z)/g_s}$, where $\phi(z)$ is the
Kodaira-Spencer field,
 with $\partial \phi$ being the effective one form
on the Riemann surface, which in our case in the classical limit is given by
\begin{equation}
\partial \phi(z)|_{class.}=\frac{\mu}{z} \ .
\end{equation}
Consequently, acting with the operator $\hat{x}$ on the classical
part of the fermion/brane at a fixed point y in the $y$-patch has 
to give
\begin{equation}
\label{Condi22}
\left<\psi(y) \right|\hat{x}(y) \left| \psi(y) \right> |_{class.}
\to e^{i \phi(y)/g_s}\hat{x}(y) e^{-i \phi(y)/g_s}=x(y)  =
\mu^2/y \ ,
\end{equation}
where $\phi_{class.}(y)=\mu
\log{y}$. 

The operator that satisfies (\ref{Condi22}) is given by
\begin{equation}
\label{XRep22}
\hat{x}(y)=-g_s^2\left[\partial_y+y\partial_y^2 \right] \ .
\end{equation}
We want to note that the above is a solution to the string equation
of the 0A string for $q=0$ found in \cite{Yin:2003iv}
\begin{equation}
\left[L_+,L_-\right]=2iM \; , \;
\left\{L_+,L_-\right\}=2M^2-\frac{1}{2} \ ,
\end{equation}
where $L_{\pm}$ are the Lax operators and $M$ is the Orlov-Shulman operator.
To see this we have to identify $L_+\to \hat{x},L_- \to y$ and 
$M=\mu=H$, in the unperturbed case, where $H$ denotes the Hamiltonian.
In addition we can also read this operator from the Ward identities
(\ref{Ws1}) which are equivalent to (\ref{Ws22}).
 
We can
use the representation of the operator $\hat{x}$ in the $y$-patch
to derive the canonical transformation which transforms the branes/fermions
from one patch to the other.
The transformation rule is quite generally
\begin{equation}
\label{Trafo1}
\psi(y)=\int dx F(x,y) \psi(x) \ .
\end{equation}
Eq. (\ref{XRep22})
then leads to the following differential equation for the kernel $F(x,y)$
\begin{equation}
\left(\partial_y+y \partial_y^2\right)F(x,y)=\frac{x}{g_s^2} F(x,y) \ .
\end{equation}
Changing variables to $z=2 \frac{\sqrt{xy}}{g_s}$ we find
\begin{equation}
z \partial_z^2F(z)+\partial_zF(z)=zF(z) \ ,
\end{equation}
which is solved by the modified Bessel function of the second kind
\begin{equation}
\label{MBF1}
F(z)=K_0(z)=K_0(2 \frac{\sqrt{xy}}{g_s}) \ .
\end{equation}
Note that (\ref{MBF1}) has the integral representation
\begin{equation}
\int_0^{\infty}dz e^{-2\frac{\sqrt{xy}}{g_s}\cosh{z}} \ ,
\end{equation}
which can be rewritten as
\begin{equation}
\int_0^{\infty}\frac{dz}{z}e^{-\frac{1}{g_s}\left(1/z+xyz\right)} \ .
\label{F}
\end{equation}
With this transformation at hand we can employ the strategy of \cite{Aganagic:2003qj},
which we will discuss in section \ref{section6.6}, to solve the corresponding B-model on 
(\ref{Geomy1})
leading to the expressions (\ref{DetA1}),(\ref{MI1}). 

In the next section we will use the brane/fermion
picture to derive the $0$A reflection coefficients
using the above transformation properties of the
fermions.

\subsection{Type 0A Reflection Coefficients}

In this section we will show that we can derive the reflection coefficients for
the $0$A strings at $R=1$ from the topological brane picture following a 
similar derivation for the $c=1$ string at self-dual radius in \cite{Aganagic:2003qj}.
We will set $g_s=1$ in this section in order to match the matrix model conventions.

For the derivation we will parameterize the Riemann surface by the coordinates $x,y$, which,
as we have seen from the reduction of the holomorphic 3-form,  
are not canonically conjugated.
We have found that for the case $q=0$  
the classical part of the KS field in the $x$-patch is given by
\begin{equation}
\partial \phi(x)=\frac{\mu}{x} \ ,
\end{equation}
such that the classical piece of the fermions/branes is given by
\begin{equation}
\psi_{cl}(x)=e^{-i\phi_{cl}}=x^{-i\mu} \ .
\end{equation}
If we evaluate the brane two point function in the $x$-patch we get
\begin{equation}
\label{XP1}
\left<0\right| \psi(\tilde{x}) \psi^*(x) \left|V\right> =
\frac{\tilde{x}^{-i \mu} x^{i\mu }}{x-\tilde{x}} \ .
\end{equation}
On the other hand the correlator of two branes in the
two respective asymptotic regions is given by
\begin{equation}
\label{XYP1}
\left<0\right| \tilde{\psi}(y) \psi^*(x) \left|V\right> =
\left(xy\right)^{i\mu}\sum_{n \ge 0} \left(xy\right)^{-n-1} R_{n+1/2} \ ,
\end{equation}
where $R_{n+1/2}$ is the reflection coefficient and
where the classical piece of the KS field
in the $y$-patch reads
\begin{equation}
\partial \phi(y)=-\frac{\mu}{y} \ ,
\end{equation}
which follows from the reduction of the 3-form.

By translating expression (\ref{XP1}) to the $y$-patch
using the transformation (\ref{Trafo1}) and comparing to (\ref{XYP1})
we can derive the reflection coefficients. We have
\begin{equation}
  \left<0\right| \tilde{\psi}(y) \psi^*(x) \left|V\right> =
\int d\tilde{x} \int dz e^{-\log{z} -\frac{1}{z} -\tilde{x}yz}
\left<0\right| \psi(\tilde{x}) \psi^*(x) \left|V\right> \ .
\end{equation}
Inserting (\ref{XP1}) and comparing to (\ref{XYP1})
we get
\begin{equation}
R_{n+1/2}=\int dz e^{-\log{z}-\frac{1}{z}} \int dx e^{-xz} x^{n- i \mu } \ .
\end{equation}
Evaluating the integrals on the positive real axis leads to the correct expression for the
perturbative reflection coefficient (\ref{ReflCoeff1})
\begin{equation}
R_{n+1/2}\sim\Gamma\left(n-i \mu +1 \right)
\Gamma\left(n-i \mu +1 \right) \ .
\end{equation}

\subsection{Ramond-Ramond Charge $q$}

In the following we will 
point out some intriguing features of the model with
$q \ne 0$. 

Following the above analysis we would in this case expect the 
integrable structure of the $0$A string to describe the
deformations of the Riemann surface
\begin{equation}
\label{RS44}
H(x,y)=xy-\mu^2-q^2/4=0 \ .
\end{equation}
The holomorphic 3-form can  be written as
\begin{equation}
\Omega=\frac{1}{2} \frac{dx \wedge dy}{\sqrt{xy-q^2/4}}\wedge\frac{dv}{v} \ .
\end{equation}
If we now do the integral over the 3-cycle we encounter a branch
cut corresponding to the two 3-cycles whose mutual separation is
determined by the parameter $q$. Depending on the choice of the
sheet the integral $\int \Omega$ over the compact 3-cycles gives the
deformation parameters $\mu \pm iq/2$. However, this
implies that we cannot simply reduce the integral of $\Omega$ over
the (non)compact 3-cycles to a 1-form on the Riemann surface
(\ref{RS44}). In addition, we have a contribution from the
singular Riemann surface $xy=0$. It would be interesting
to understand this feature better by investigating in detail
the disk partition function for the non-compact
branes fibered over the Riemann surfaces \cite{Aganagic:2000gs}.

When we take into account the second boundary
contribution by using effective branes/fermions on the Riemann surface with disk partition function
$\phi_{class.}(y)=\left(\mu+iq/2\right)\log{y}$ we can run the machinery of the
$q=0$ case. In particular, we can use the argument in (\ref{Condi22}) for the
Riemann surface (\ref{RS44}) to find the operator
\begin{equation}
\label{xDiff22}
\hat{x}(y)=-g_s^2\left[(1+q)\partial_y+y\partial_y^2 \right] \ .
\end{equation}
We have rescaled $q$ by a factor of $g_s$, $\phi_{class.}(y)=\left(\mu+iq g_s/2\right)
\log{y}$. This dependence on $g_s$ may hint that while $\mu$ is related to the complex deformation due to 
the backreaction of $N \sim 1/g_s$
compact branes, $q$ is rather related to the non-compact branes in some novel sense.

As for the case $q=0$, the above is a solution to the string equation
of the 0A string \cite{Yin:2003iv}, this time with finite q,
\begin{equation}
\left[L_+,L_-\right]=2iM \; , \;
\left\{L_+,L_-\right\}=2M^2+\frac{q^2-1}{2} \ .
\end{equation}
Indeed, (\ref{xDiff22}) is also the operator that we read from the Ward identities
(\ref{Ws1}).

\subsection{B-branes and Kontsevich-like Matrix Models}
\label{section6.6}

In the following we will use  \cite{Aganagic:2003qj} to
 interpret the  Kontsevich-like matrix model,
derived in 
section 5,
as the theory on the worldvolume of non-compact B-branes.
We consider the case $q=0$. 

Consider a Calabi-Yau space of the form (\ref{uw}),
as a $\bC$ fibration over a two-dimensional complex
plane spanned by $x,y$.
The non-compact B-branes 
wrap the fiber directions and intersect the Riemann surface 
$H(x,y)=0$ at points $x_i$. Denote
\begin{equation}
\Lambda={\rm diag}(x_1,\cdots, x_N) \ .
\end{equation}
The non-compact branes backreact on the geometry and deform the complex structure
with deformation parameters $t_n$ given by 
\begin{equation}
t_n={g_s\over n}\tr\Lambda^{-n} \ .
\end{equation}
This can be seen by
realizing the non-compact branes at points $x_i$  by fermions (\ref{fermion})
and computing the one-point function of the Kodaira-Spencer field $\phi(x)$ in the background
of the fermions.

We consider the two patches $x=\infty$ and $y=\infty$.
The deformations of the complex structure in the $x$ and $y$ patch are denoted by
$\bar{t}_n$ and $t_n$ respectively, and are the couplings to positive and negative
momentum tachyons.
Let us set $t_n=0$, which implies 
that $\partial/\partial\bar{t}_n=0$, since
all the tachyon scattering amplitudes vanish by momentum
conservation.
Thus, the   Kodaira-Spencer field takes the form 
\begin{equation}
y=\partial\phi(x)= \mu x^{-1}-\sum_{n>0}n\bar{t}_nx^{n-1} \ .
\label{KS:x}
\end{equation}
In order to deform the complex structure  in the $y$ patch,
we describe the non-compact B-branes in the $x$ patch and translate
them to the $y$ patch by using the canonical transform 
with the fermions transforming from patch to patch by (\ref{cano:fermion}).

In the $x$ patch the non-compact B-branes can be described by $\psi(x)=e^{-i\phi(x)/g_s}$.
Their $N$-point function can
be evaluated by using (\ref{KS:x}) as
\begin{equation}
\langle \psi(x_1)\cdots\psi(x_N) \rangle=\Delta(x)\, e^{-\sum \phi(x_i)}
=\Delta(x)\,e^{\sum_i\left(
-{i\mu\over g_s}\log x_i+{i\over g_s}\sum_n \bar{t}_nx^n_i\right)} \ ,
\end{equation}
where $\Delta(x)=\prod_{i<j}(x_i-x_j)$.
By performing the canonical transform (\ref{Trafo1}) with (\ref{F}), we get the Kontsevich-like integral
of section 5.

\section{Quiver Gauge Theories and Matrix Models}

We will consider D-branes wrapping holomorphic cycles in Calabi-Yau
3-folds. In general, integrating out the open string degrees of freedom 
results in a deformed geometry, which is interpreted as the backreaction
of the D-branes on the original geometry.
In the topological string framework the back reaction affects the closed strings
by changing the periods of the holomorphic 3-form:
The Calabi-Yau space undergoes a transition where holomorphic cycles
shrink and 3-cycles are opened up.
In the language of four-dimensional ${\cal N}=1$ supersymmetric gauge theory,
the D-branes wrapping in the resolved geometry provide the UV description,
while the IR physics is described by the deformed geometry after the transition.

For confining gauge theories, one assumes that the relevant IR degrees of freedom
are the glueball superfields $S_i$. In the deformed geometry,
the glueball superfields are related to the integrals of $\Omega$ by
\beq
S_i = \int_{3-cycle} \Omega \ .
\eeq
The partition function of the topological field theory, as a function of the deformation 
parameters, computes the holomorphic F-terms of the gauge theory as a function
of the glueball superfields.
This has also a matrix model description 
\cite{Dijkgraaf:2002fc,Dijkgraaf:2002vw,Dijkgraaf:2002dh,Dijkgraaf:2003xk}.
One can extend the discussion to  
 the ring of chiral operators  of
the supersymmetric gauge theory and its correlators, and relate it
to the matrix model and topological field theory pictures.

The whole picture can sometimes be realized by a non-critical string theory.
In the case of the bosonic $c=1$ string at the self-dual radius, the 
corresponding quiver gauge theory and matrix model are described 
by the ${\hat A}_1$ diagram.
In this section we will construct the quiver gauge theories and matrix model pictures of
the $\hat{c}=1$ strings at the radii (\ref{topp}).

\subsection{Super Affine 0A} 

We will start by reviewing several aspects of the relevant resolved and deformed Calabi-Yau
geometries. We will then construct the quiver gauge theory and matrix model corresponding
to the superaffine 0A strings.

\subsubsection{The Geometry}

We have seen in section 3 that 
the ground ring of the super-affine 0A string at the self-dual
radius is a $Z_2$ quotient of the conifold (\ref{z22}). 
We will need some details on the resolution and deformation of this
singular space.
In this case, it is convenient to use toric variety techniques. 

Here, the toric data is a set of two-vectors,
\begin{eqnarray}\label{twovecsa}
  w_1=(1,0),\quad w_2=(-1,0),\quad w_3=(0,1),\quad w_4=(0,-1),\quad w_5=(0,0) \ .
\end{eqnarray}
As necessary for this class of (Gorenstein canonical)
singularities, these two-vectors lie in the interior or boundary of a polygon in 
${\mathbb R}^2$.

{}From the vectors (\ref{twovecsa}) one computes integer charges $Q^i_l$ satisfying, 
\begin{eqnarray}
  \sum_{i=1}^k Q^i_lw_i=0,\quad \sum_{i=1}^k Q^i_l=0 \ .
\end{eqnarray}
In our case we have
\begin{eqnarray}
&&\;\,\begin{array}{ccccc}z_1&z_2&z_3&z_4&z_5\end{array}\\
  \left(\begin{array}{c}Q_1\\ Q_2\end{array}\right)&=&\left(\begin{array}{ccccc}\,1&\,1&\,0&\,0&-2\\\,0&\,0&\,1&\,1&-2\end{array}\right) \ .
\end{eqnarray}

These charges represent a $U(1)^2$  action on the coordinates $z_i \in
\mathbb C^5$.
The singular space is the symplectic quotient of  $\mathbb C^5$ by  $U(1)^2$ defined by the (D-term) equations
 \begin{eqnarray}\label{ressp1}
  |z_1|^2+|z_2|^2-2|z_5|^2&=&0 \ ,\nn
  |z_3|^2+|z_4|^2-2|z_5|^2&=&0 \ ,
\end{eqnarray}
modulo the $U(1)^2$  action.
The resolved space is obtained by introducing two (FI) parameters  $t_1,t_2$ , such that
\begin{eqnarray}\label{ressp2}
  |z_1|^2+|z_2|^2-2|z_5|^2&=&t_1 \ ,\nn
  |z_3|^2+|z_4|^2-2|z_5|^2&=&t_2 \ .
\end{eqnarray}
We can use the second equation of (\ref{ressp2}) to eliminate
the coordinate $z_5$ up to a phase factor, which can be absorbed 
by the second $U(1)$.
We are left with a residual $Z_2$ symmetry which leaves $z_5$ invariant,
and acts as $(z_3,z_4)\rightarrow -(z_3,z_4)$.
Thus, we can view the space as a $Z_2$ orbifold of the resolved conifold
\begin{eqnarray}\label{s0aM}
 && |z_1|^2+|z_2|^2-|z_3|^2-|z_4|^2=t,\quad t=t_1-t_2,\nn
 && Z_2: (z_3,z_4)\rightarrow -(z_3,z_4)  \ ,
\end{eqnarray}
where the first equation describes the resolved conifold.
The resolved conifold has the structure $O(-1)\oplus O(-1) \rightarrow
\mathbb{P}^1$.
For $t>0$, the $Z_2$ 
acts on the fiber leaving the  $\mathbb{P}^1$ fixed.
It can be viewed also as an $A_1$ singularity fibered over a $\mathbb{P}^1$ with volume $t_1-t_2$.
This description is valid in the semi classical regime of the  K\"{a}hler moduli space 
$t_1,t_2 \rightarrow -\infty$
with $t_1-t_2$ finite.

There is another semi classical regime of interest defined by $t_1,t_2 \rightarrow \infty$.
In this regime equations (\ref{ressp2}) describe a bundle over  the Hirzebruch surface 
$\mathbb{F}_0 = \mathbb{P}^1\times \mathbb{P}^1$ with sizes $t_1$ and $t_2$ of the two $\mathbb{P}^1$'s.

In order to see that indeed the space we are considering is the 3-fold described by the
ground ring of the super-affine 0A string at the self-dual radius we will review now
an alternative 
description of the toric singularity.
This is a description in terms of a 
collection of polynomial equations, written in terms 
of $U(1)^2$-invariant monomials of the coordinates $z_i$. 
The monomials are generated by the nine elements
\begin{eqnarray}
  B_{ij}=\left(\begin{array}{ccc}z_1^2z_3^2z_5&z_1^2z_3z_4z_5&z_1^2z_4^2z_5\\
                               z_1z_2z_3^2z_5&z_1z_2z_3z_4z_5&z_1z_2z_4^2z_5\\
                               z_2^2z_3^2z_5&z_2^2z_3z_4z_5&z_2^2z_4^2z_5
                               \end{array}\right) \ .
\end{eqnarray}
This matrix can be identified with the matrix $b_{ij}$ in (\ref{ringelements}). 
The monomials $B_{ij}$ satisfy the same relations as $b_{ij}$.
For instance, 
\begin{eqnarray}
  B_{11}B_{22}=(z_1^2z_3^2z_5)\,(z_1z_2z_3z_4z_5)=(z_1^2z_3z_4z_5)(z_1z_2z_3^2z_5)=B_{12}B_{21} \ .
\end{eqnarray}

We can identify in the toric picture
the $sl(2)_L\times sl(2)_R$ affine Lie algebra of super affine 0A string theory. 
The $sl(2)_L$ algebra is represented by the differential operators 
\begin{eqnarray}
  L_0=z_1\partial_1-z_2\partial_1,\quad L_-=z_1\partial_2,\quad L_+=z_2\partial_1 \ .
\end{eqnarray}
The $sl(2)_R$ is realized by similar operators with
the replacement $(1\rightarrow 3,2\rightarrow 4)$.

%There is a  unique $sl(2)_L\times sl(2)_R$ deformation of the singular complex 
%space described by the $B_{ij}$ elements and their relations.
The relations among the $B_{ij}$ can be organized in $sl(2)_L\times sl(2)_R$
representations. The unique $sl(2)_L\times sl(2)_R$ invariant
deformation is given by changing only the singlet relation to
\begin{eqnarray}
 2B_{22}B_{22}-2(B_{32}B_{12}+B_{23}B_{21})+B_{31}B_{13}+B_{11}B_{33}= \mu^2 \ . 
\end{eqnarray}
Then the deformed space has the structure $T^*(S^3/Z_2)$, where the $Z_2$ acts freely on $S^3$.

\subsubsection{Quiver Gauge Theories}

The UV description of the quiver gauge theories is found by D-branes wrapping in the
resolved geometry.
Consider the semi classical regime, where the resolved geometry is described by the $Z_2$
quotient of the resolved conifold
(\ref{s0aM}).
As a starting point consider $N_0$ D3-branes placed at the singularity $t=0$.
This system can be analysed by taking $2N_0$ D3-branes \cite{Klebanov:1998hh} at the conifold and performing
the $Z_2$ quotient \cite{Douglas:1996sw}. One gets \cite{Morrison:1998cs}
the gauge group $SU(N_0)^4$ and  matter fields 
in bifundamental representations.

One can add fractional $N_0$ D3-branes by wrapping $N_1$
D5-branes on the resolved $\mathbb{P}^1$ in (\ref{s0aM}), or by
wrapping $2N_1$ D5-branes on the resolved conifold
\cite{Klebanov:2000nc,Klebanov:2000hb}
and performing
the $Z_2$ quotient.
To see how the quotient works, let us start with the $\hat{A}_1$ quiver
gauge theory with the matter content 
\begin{eqnarray}
  \begin{array}{ccc}&SU(2N_0)&SU(2N_0+2N_1)\\
    \bA_i&\Box&\bar\Box\\
    \bB_j&\bar\Box&\Box\\
    \bPhi^+&{\rm \bf adj.}&\\
    \bPhi^-&&{\rm \bf adj.}
  \end{array}
\label{original:a1}
\end{eqnarray}
and the superpotential given by
\begin{equation}
W={1\over 2}\left(\tr\, \bPhi^{+2}-\tr\, \bPhi^{-2}\right)
-\tr\left(\bA_i\bPhi^+\bB_i\right)
-\tr\left(\bA_i\bPhi^-\bB_i\right) \, .
\label{wtree:a1}
\end{equation}
The quotient acts on the fields as
\begin{equation}
\bA_i\rightarrow\bA_i,~~
\bB_i\rightarrow -\bB_i,~~
\bPhi_\pm\rightarrow -\bPhi_\pm \, ,
\end{equation}
which is accompanied by the action on the Chan-Paton factors as
\begin{equation}
\gamma_{2N_0}={\rm diag}(I_{N_0},-I_{N_0}),~~
\gamma_{2N_0+2N_1}={\rm diag}(I_{N_0+N_1},-I_{N_0+N_1}).
\end{equation}
Note that the $\bZ_2$ is a symmetry of the $\hat{A}_1$ theory.
The quotient thus takes the form
\begin{eqnarray}
&&\gamma_{2N_0}\bA_i\gamma_{2N_0+2N_1}=+\bA_i,~~
\gamma_{2N_0+2N_1}\bB_i\gamma_{2N_0}=-\bB_i,~~\nn
&&\gamma_{2N_0}\bPhi^+\gamma_{2N_0}=-\bPhi^+,~~
\gamma_{2N_0+2N_1}\bPhi^-\gamma_{2N_0+2N_1}=-\bPhi^- \, .
\end{eqnarray}
The matter content of the quotient of the ${\hat A}_1$ quiver theory
becomes
\begin{eqnarray}
  \begin{array}{ccccc}&SU(N_0)&SU(N_0)&SU(N_0+N_1)&SU(N_0+N_1)\\
    A_i&\Box&&\bar\Box&\\
    \tilde{A}_i&&\Box&&\bar\Box\\
    B_j&\bar\Box&&&\Box\\
    \tilde{B}_j&&\bar\Box&\Box&\\
    \Phi^+_1&\Box&\bar\Box&&\\
    \Phi^+_2&\bar\Box&\Box&&\\
    \Phi^-_1&&&\Box&\bar\Box\\
    \Phi^-_2&&&\bar\Box&\Box\end{array}
\end{eqnarray}
with $i=1,2$.
$A_i,\tilde A_i$ and $B_j,\tilde B_j$ transform as doublets under 
$sl(2)_L$ and $sl(2)_R$ respectively. 
%$\Phi_{1,2}^{\pm}$ arise
%from the adjoints in the $\hat A_1$ quiver and are massive. 
%The $sl(2)_L\times sl(2)_R$ symmetry appears after integrating out the fields $\Phi_{1,2}^{\pm}$. 
The superpotential can be obtained from (\ref{wtree:a1}) by restricting
to the fields which survive the projection
\begin{eqnarray}
  W_{tree}&=&W_{0}+W_{1}\nn
  W_{0}&=&m\, tr(\Phi^+_1\Phi^+_2)-m\, tr(\Phi^-_1\Phi^-_2) \ , \nn
  W_{1}&=&-tr(\tilde A_i\Phi^+_1\tilde B_i)-tr(A_i\Phi^+_2B_i)-
tr(\tilde B_i\Phi^-_1 A_i)-tr(B_i\Phi^-_2\tilde A_i) \ .
\label{sup}
\end{eqnarray}
Integrating out the massive fields one gets the quartic superpotential
\begin{eqnarray}
  W =  \varepsilon^{ik}\varepsilon^{jl}\,tr(A_j\tilde B_k\tilde A_lB_i) \ .
%\label{suppot}
\label{w:s0a}
\end{eqnarray}

\begin{figure}[htb]
\begin{center}
\epsfxsize=3in\leavevmode\epsfbox{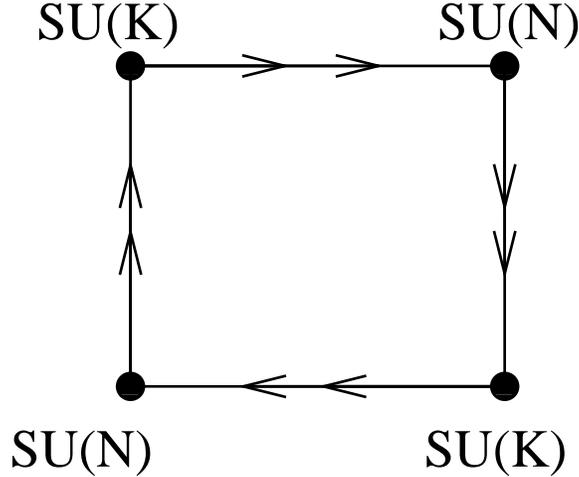}
\end{center}
\caption{Quiver gauge theory
for the $A_1$ fibered over ${\bf P}^1$ geometry, $N=N_0$, $K=N_0-N_1$.}
\label{quive1}
\end{figure}

As in the original $\hat{A}_1$ quiver gauge theory, the quotient gauge
theory exhibits RG cascades.
Suppose that the $SU(N_0+N_1)\times SU(N_0+N_1)$ gauge groups are
strongly coupled.
One can apply a Seiberg duality for the two gauge groups simultaneously
and find a weakly coupled description. The dual theory
is
\begin{eqnarray}
  \begin{array}{ccccc}&SU(N_0-N_1)&SU(N_0-N_1)&SU(N_0)&SU(N_0)\\
    a_i&\Box&&\bar\Box&\\
    \tilde{a}_i&&\Box&&\bar\Box\\
    b_j&\bar\Box&&&\Box\\
    \tilde{b}_j&&\bar\Box&\Box&\\
    M_{ij}&&&\Box&\bar\Box\\
    \tilde{M}_{ij}&&&\bar\Box&\Box\\
%    \Phi^+_1&\Box&\bar\Box&&\\
%    \Phi^+_2&\bar\Box&\Box&&\\
%    \Phi^-_1&&&\Box&\bar\Box\\
%    \Phi^-_2&&&\bar\Box&\Box
   \end{array}
\label{quiv3}
\end{eqnarray}
with the superpotential 
\begin{equation}
W=\varepsilon^{ik}\varepsilon^{jl}\tr( M_{ik}\tilde{M}_{jl}) 
+\tr\left(a_iM_{ij}b_j+\tilde{a}_i\tilde{M}_{ij}\tilde{b}_j\right) \, .
\end{equation}
Here
\begin{equation}
M_{ij}=A_i\tilde{B}_j,~~
\tilde{M}_{ij}=\tilde{A}_iB_j \, ,
\end{equation}
and $a_i,\tilde{a}_i,b_j,\tilde{b}_j$ are dual quarks.
By integrating out the meson fields, one gets the superpotential
\begin{eqnarray}
  W =  \varepsilon^{ik}\varepsilon^{jl}\,\tr(a_i\tilde b_j\tilde a_kb_l) \ .
\end{eqnarray}
This is identical to (\ref{w:s0a}).
The gauge groups and matter content are encoded in the
quiver diagram, see figure \ref{quive1}.

Let us now discuss
the classical moduli space of vacua of the quiver theory
given in figure \ref{quive1}, 
which is parametrized by the invariant traces of the chiral
operators. 
Contrary to the notation in (\ref{quiv3}), 
the bifundamental fields $a,\tilde{a},b,\tilde{b}$ will be denoted by
the capital letters $A,\tilde{A},B,\tilde{B}$ from now on.
The generators of the chiral ring are quartic
expressions in the fields of the form 
\begin{eqnarray}
 O_{i_1j_1i_2j_2} = A_{i_1}\tilde B_{j_1}\tilde A_{i_2}B_{j_2} \ . 
\end{eqnarray}
 These operators are symmetric in the indices $i_l$ as well as in the indices $j_l$, as can 
be seen from the field equations of (\ref{suppot}), 
\begin{eqnarray}\label{eoms}
  A_{i_1}\tilde B_{j_1}\tilde A_{i_2}=A_{i_2}\tilde B_{j_1}\tilde A_{i_1},\quad \tilde B_{j_1}\tilde A_{i_2}B_{j_2}=\tilde B_{j_2}\tilde A_{i_2}B_{j_1} \ .
\end{eqnarray}
$A_{i_1}\tilde B_{j_1}\tilde A_{i_2}B_{j_2}$ can be 
matched with the  super-affine 0A ground ring generators $b_{ij}$ (\ref{ringelements}),
by noting that one can write the quadratic
expressions (\ref{eoms}) as $a_{i_1j_1}a_{i_2j_2}$ subject to the relation (\ref{det}).
One can verify the ring relations by using the field equations. For instance,
\begin{eqnarray}
  b_{11}b_{22}=O_{1111}O_{1122}=O_{1112}O_{1121}=b_{12}b_{21} \ .
\end{eqnarray}

Let us consider now the IR dynamics of the 
quiver gauge theory. 
Of particular interest for us is the case
of $N_0$ being an integer multiple of $N_1$.
In this case, after the duality cascades, one ends up with a
confining gauge theory with 
%$N_0=N_1=N$, which we will consider first. In this case 
the gauge group $SU(N_1) \times SU(N_1)$ 
%with no massless matter fields and 
and the massive bifundamentals
$\Phi_i^+, i=1,2$.
In the deformed geometry $N=N_1$ 
sets the size of $S^3$ which we identify with the
glueball superfield $S$.
The holomorphic F-terms of this theory ${\cal F}(S)$ as a function of  $S$ are twice that
of  $SU(N)$ SYM and are related to the perturbative super affine 0A free energy
as\footnote{Of course we have to take into account possible non-trivial relations
between the free energy and the holomorphic F-terms. Recall, for instance, that the planar
free energy is related to the superpotential by a derivative w.r.t. S 
\cite{Dijkgraaf:2002dh}.}
\beq
{\cal F}_{\rm SYM}(S) = \cF_{\rm 0A}^{\rm super-aff.}(R_{\rm self-dual})(\mu) \ ,
\eeq
where $S\sim \mu$.

%When $N_1 \neq N_2$ some of the gauge couplings become strong and one has to run the a cascade
%of Seiberg dualities in order to study the IR physics.
%This can be analysed in detail, but will not be done here.
%We should make remark though: We expect that the quivers corresponding to type 0A strings
%to have the same IR dynamics as the for $N_0=N_1$.

The quiver gauge theory arising from D-branes wrapping
in the semi classical regime $t_1,t_2 \rightarrow \infty$
is described by figure 2 and was argued to be (toric) Seiberg dual to the quiver theory
of (\ref{quiv3}) \cite{Beasley:2001zp}\cite{Feng:2001bn}.

\begin{figure}[htb]
\begin{center}
\epsfxsize=3in\leavevmode\epsfbox{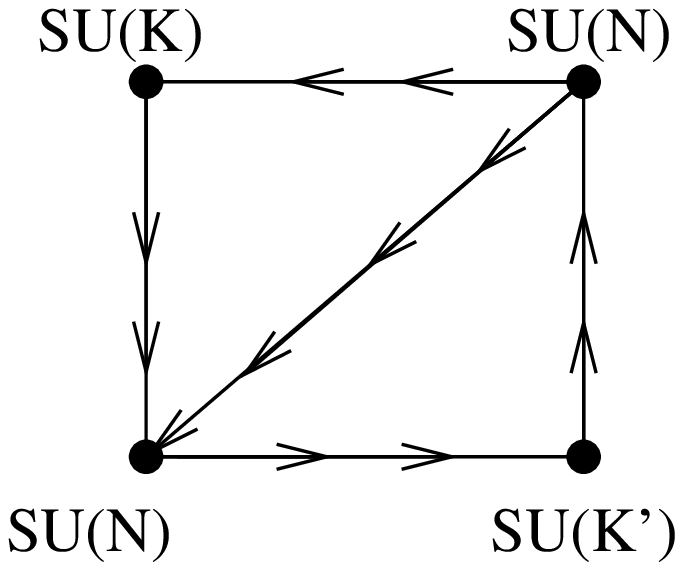}
\end{center}
\caption{Quiver
gauge theory for the bundle over ${\bf P}^1\times {\bf P}^1$ geometry
$N=N_0$, $K=N_0-N_1$,  $K'=N_0+N_1$.}
\label{quive2}
\end{figure} 
\subsubsection{Matrix Model} 

One can also write a 
DV matrix integral description of the quiver gauge theory F-terms.
It takes the form:
\beq
Z = {1\over V}\int d\Phi_i^+ d\Phi_i^- dA_i dB_j d\tilde{A}_i d\tilde{B}_j exp
\left[-{1\over g_s} W_{tree}(\Phi_i^+, \Phi_i^- A_i,B_j, \tilde{A}_i, \tilde{B}_j)\right] \ ,
\eeq
with  $W_{tree}$ given by (\ref{sup}), and $V$ is the volume of the groups
$SU(\hat{N_0})^2\times SU(\hat{N_0}-\hat{N_1})^2$.

For $N_0=N_1=N$ one can use the matrix integral 
\begin{eqnarray}
 Z=\frac{1}{V(SU(\hat{N})\times SU(\hat{N}))}\int d\Phi^{+}_{1}d\Phi^{+}_{2} e^{-\frac{m}{g_s}\,tr(\Phi^+_1\Phi^+_2)}
\ ,
\end{eqnarray}
which gives perturbatively
\beq
 {\cal F}_{\rm matrix~ model}(S) = 2\,\times\,{\cal F}_{c=1}(R_{\rm self-dual})(\mu) 
= {\cal F}_{\rm 0A}^{\rm super-affine}(R_{\rm self-dual})(\mu) \ ,
\eeq
with $S=g_s \hat{N}$.
Thus, we see that
the F-terms of the gauge theory in figure 1 with $N_0=N_1=N$ 
are related to the partition function of superaffine 0A string at the radius (\ref{topp}).

\subsection{Circle Line Theories} 

We will start by reviewing several aspects of the relevant resolved and deformed Calabi-Yau
geometries. We will then construct the quiver gauge theory and matrix model corresponding
to the 0A (0B) strings.

\subsubsection{The Geometry}

The $\mu =0$
ground ring of the  0A string at $R=1$ and of the 0B string at $R=2$ 
define the same singular space.
Here, the toric data is a set of two-vectors,
\begin{eqnarray}\label{twovec0a}
  w_1=(2,0),\quad w_2=(2,1),\quad w_3=(1,0),\quad w_4=(1,1),\quad w_5=(0,0),\quad w_6=(0,1) 
\ .
\end{eqnarray}

{}From the vectors (\ref{twovec0a}) one computes integer charges $Q^i_l$ 
\begin{eqnarray}
&&\;\,\begin{array}{cccccc}\,\,\,z_1&\,\,z_2&\,\,z_3&\,\,z_4&\,\,z_5&\,\,z_6\end{array} \nonumber \\
  \left(\begin{array}{c}Q_1\\ Q_2\\Q_3\end{array}\right)&=&\left(\begin{array}{cccccc}\,1&\,-1&\,-1&\,1&\,0&\,0\\\,0&\,1&\,0&-2&0&1\\\,0&\,0&\,1&-1&\,-1&1\end{array}\right) \ .
\end{eqnarray}
The
resolved singularity is described by the equations 
\begin{eqnarray}\label{ressp0a}
  |z_1|^2-|z_2|^2-|z_3|^2+|z_4|^2&=&t_1,\nn
  |z_2|^2-2|z_4|^2+|z_6|^2&=&t_2,\nn
  |z_3|^2-|z_4|^2-|z_5|^2+|z_6|^2&=&t_3.
\end{eqnarray}
modulo the $U(1)^3$ action.
For zero resolution parameters $t_1,t_2,t_3$ these equations define the toric singularity 
of 0A at $R=1$.

In order to see that indeed the space we are considering is the 3-fold described by the
ground ring of the 0A string at the radius (\ref{topp}),
we check that the ring
of $U(1)^3$-invariant monomials of the coordinates $z_i$,
is generated by the four elements
\beq
u= z_1z_3z_5,~~~v= z_2z_4z_6,~~~x=z_1^2z_2^2z_3z_4~~~,y=z_3z_4z_5^2z_6^2 \ ,
\eeq
with the relation
\beq\label{gr0Atoric}
(uv)^2 - xy = 0 \ ,
\eeq
which is the ground ring relation (\ref{gr0A}).

Consider the semi classical regime $t_2 \rightarrow -\infty$.
The second equation of (\ref{ressp0a}) eliminates $z_4$
up to a phase factor, which can be fixed
by using the second $U(1)$ symmetry. 
This leaves a $Z_2$ 
action $(z_2,z_6)\rightarrow (-z_2,-z_6)$. 
Consider in addition the regime  $t_1 \rightarrow -\infty, t_3 \rightarrow \infty$,
with $t_1+t_3=t$, with $t$ fixed.
Adding the first and third equations (\ref{ressp0a}) we get
\begin{eqnarray}\label{quot}
  &&|z_1|^2+|z_6|^2-|z_2|^2-|z_5|^2=t \ ,\\
  &&Z_2:\;\;(z_2,z_6)\rightarrow(-z_2,-z_6) \ .
\end{eqnarray}
This is the the resolved conifold $O(-1)\oplus O(-1) \rightarrow  \mathbb{P}^1$, and the $Z_2$ 
acts on both the fiber and the  $\mathbb{P}^1$.

\subsubsection{Type 0A Quiver Gauge Theories}

One can construct the quiver gauge theory by wrapping branes
on the resolved singularity in the semi classical regime discussed above,
as in \cite{Cachazo:2001sg}.
Alternatively, we can start from the $\hat{A}_1$ quiver gauge theory
(\ref{original:a1}) and take a $Z_2$ quotient defined by
\begin{eqnarray}
\bZ_2:~~~~  \bA_2\rightarrow -\bA_2,\quad \bB_1\rightarrow -\bB_1,\quad 
\bPhi^{\pm}\rightarrow-\bPhi^{\pm} \, ,
\end{eqnarray}
together with the action on the Chan-Paton factors.
One can see that the matter content is given by
\begin{eqnarray}
  \begin{array}{ccccc}&SU(N_0+N_1)&SU(N_0)&SU(N_0+N_1)&SU(N_0)\\
    \tilde{A}_1&\bar\Box&\Box&&\\
    \tilde{B}_2&\Box&\bar\Box&&\\
    \tilde{A}_2&&\Box&\bar\Box&\\
    \tilde{B}_1&&\bar\Box&\Box&\\
    A_1&&&\bar\Box&\Box\\
    B_2&&&\Box&\bar\Box\\
    A_2&\bar\Box&&&\Box\\
    B_1&\Box&&&\bar\Box\\
    \Phi^+_1&\Box&&\bar\Box&\\
    \Phi^+_2&\bar\Box&&\Box&\\
    \Phi^-_1&&\Box&&\bar\Box\\
    \Phi^-_2&&\bar\Box&&\Box\end{array}  %\ .
\end{eqnarray}
%The fields $\Phi_{1,2}^{\pm}$ arise
% from the adjoints in the $\hat A_1$ quiver and are massive. 
Note that the $sl(2)_L\times sl(2)_R$ symmetry is broken. 

The superpotential reads
\begin{eqnarray}
  W_{tree}&=&W_{0}+W_{1} \ , \nn
  W_{0}&=&m\,\left[ tr(\Phi^+_1\Phi^+_2)-\, tr(\Phi^-_1\Phi^-_2)\right] \ , \nn
  W_{1}&=&-tr(A_1\Phi^+_2 B_1)-tr(A_2\Phi^+_1 B_2)-tr(\tilde A_1\Phi^+_1\tilde B_1)
-tr(\tilde A_2\Phi^+_2\tilde B_2)-
\nn
&&-tr(B_1\Phi^-_2 \tilde A_1)-tr(B_2\Phi^-_2 \tilde A_2)-tr(\tilde B_1\Phi^-_1 A_1)-tr(\tilde B_2\Phi^-_1 A_2)
\label{W0A}
\end{eqnarray}
Integrating out the massive fields  one gets
\begin{eqnarray}\label{suppot}
W&\sim&\, tr(A_1B_2 A_2B_1)+tr(\tilde A_1\tilde B_2\tilde A_2 \tilde B_1)- \nn
  &&\quad\quad -tr(A_1\tilde B_1\tilde A_2B_2)-tr(\tilde A_1B_1 A_2\tilde B_2) \ .
\end{eqnarray}

%Consider $N_0$ D3-branes and $N_1$ wrapped D5-branes.
%One has \cite{Feng:2001bn} 

\begin{figure}[htb]
\begin{center}
\epsfxsize=3in\leavevmode\epsfbox{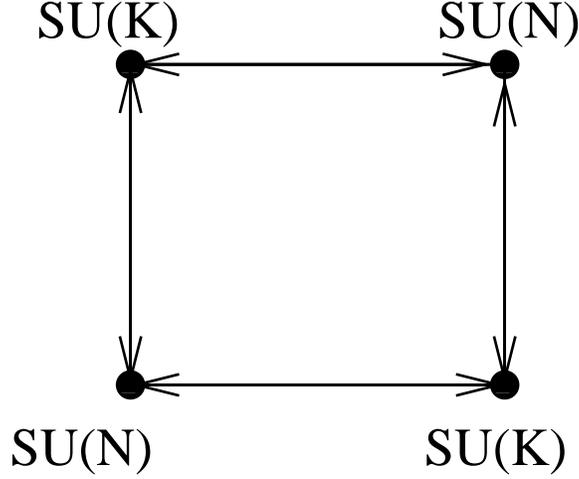}
\end{center}
\caption{Quiver gauge theory
for the 0A circle line theory at $R=1$,
$N=N_0$, $K=N_0+N_1$.}
\label{quive:0a}
\end{figure}

The information of the gauge group and matter content is encoded in
the quiver diagram, see figure \ref{quive:0a}.

The geometry (\ref{gr0Atoric}) is the classical moduli space of vacua of the $D3$ branes. 
It is parametrized by the invariant traces of the projection of operators of the form
\begin{eqnarray}
  O_{i_1j_1i_2j_2}=A_{i_1}B_{j_1}A_{i_2}B_{j_2} \ .
\end{eqnarray}
On the D3-brane branch these operators are generated by
\begin{eqnarray}
  a_{12}=\tilde A_1\tilde B_2,&&\quad a_{21}=\tilde A_2\tilde B_1,  \nonumber \\
  b_{11}=\tilde A_1 B_1 A_1\tilde B_1,&&\quad b_{33}=\tilde A_2 B_2 A_2 \tilde B_2.
\end{eqnarray}

The classical field equations read
\begin{eqnarray}\label{eoms0a}
  B_{2}A_{2}B_{1}=\tilde B_{1}\tilde A_{2}B_{2},
\quad \tilde B_{2}\tilde A_{2}\tilde B_{1}=B_{1}A_{2}\tilde B_{2} ,
  \quad B_{1}A_{1}B_{2}=\tilde B_{2}\tilde A_{1}B_{1},
\quad \tilde B_{1}\tilde A_{1}\tilde B_{2}=B_{2}A_{1}\tilde B_{1} ,
\end{eqnarray}
and 
\begin{eqnarray}\label{eoms0a2}
  A_{1}B_{2}A_{2}=A_2\tilde B_{2}\tilde A_{1},
\quad A_2B_{1}A_{1}=A_{1}\tilde B_{1}\tilde A_{2},
  \quad \tilde A_{1}\tilde B_{2}\tilde A_{2}=\tilde A_2B_{2}A_{1},
\quad \tilde A_2\tilde B_{1}\tilde A_{1}=\tilde A_1B_{1}A_{2}.
\end{eqnarray}
One can verify the ring relations by using these equations,
\begin{eqnarray}
  (a_{12}a_{21})^2=(\tilde A_1\tilde B_2\tilde A_2\tilde B_1)^2=(\tilde A_1 B_1A_1\tilde B_1 \tilde A_2 B_2 A_2 \tilde B_2)=b_{11}b_{33}.
\end{eqnarray}

Let us consider now the IR dynamics of the 
quiver gauge theory. This depends on the numbers 
$N_0$ and $N_1$, and is subject to Seiberg-like
dualities.
A particular simple case is the choice $N_0=0$ and $N_1=N$, which we will consider first.
In this case the gauge group is $SU(N) \times SU(N)$ 
with no massless matter
and with massive bifundamentals  $\Phi_i^+, i=1,2$.
This is the same theory as the one corresponding to the super affine theory.
In the deformed geometry $N$ sets the size of $S^3$ which we identify with the
glueball superfield $S$.
The holomorphic F-terms of this theory $\cal{F}(S)$ as a function of  $S$ are twice that
of  $SU(N)$ SYM and are related to the perturbative type 0A free energy
as
\beq
{\cal F}_{\rm SYM}(S) = \cF_{\rm 0A}(R=1)(\mu) \ ,
\eeq
where $S\sim \mu$.

For different $N_0,N_1$,
some of the gauge couplings become strong and one has to run a cascade
of Seiberg dualities in order to study the IR physics.
This can be analysed in detail, but will not be done here.
As we noted before,
we expect the quivers corresponding to type 0A strings
to have the same IR dynamics as the above.
 
\subsubsection{Deformed Ring Geometry}

Gluino condensation in the D5-branes worldvolume
theories deforms the ring relation (\ref{gr0A}) and (\ref{gr0Atoric}). The deformed geometry can be
seen by a D3-brane probe. The probe is described by adding the $U(1)$s to the quiver theory. 
To take into account the gauge dynamics on the D5-branes worldvolume, 
one introduces the 
Affleck-Dine-Seiberg superpotential \cite{Affleck:1983mk}, and solves 
for the meson fields $M$ and $\tilde M$ 
\begin{eqnarray}
  W_{ADS}=(N-2)\left(\frac{\Lambda^{3N-2}}{\det{M}}\right)^{\frac{1}{N-2}}+(N-2)\left(\frac{\tilde\Lambda^{3N-2}}{\det{\tilde M}}\right)^{\frac{1}{N-2}},\\
  W_{tree}={1\over m}\left(m_{22} \tilde m_{22}+m_{11} \tilde m_{11}-m_{12} m_{21}-\tilde m_{12}\tilde m_{21}\right),
\end{eqnarray}
with 
\begin{eqnarray}
  M=\left(\begin{array}{cc}\tilde A_2\tilde B_1&\tilde A_2 B_2\\A_1\tilde B_1&A_1B_2\end{array}\right),\quad \tilde M=\left(\begin{array}{cc}\tilde A_1\tilde B_2&\tilde A_1 B_1\\A_2\tilde B_2&A_2B_1\end{array}\right).
\end{eqnarray}
The equations of motion give
\begin{eqnarray}
&&  m_{11}m_{22}-m_{12}m_{21}=\left(\Lambda^{3N-2}m^{N-2}\right)^{{1\over N-1}},\quad \tilde m_{11}\tilde m_{22}-\tilde m_{12}\tilde m_{21}=\left(\tilde{\Lambda}^{3N-2}m^{N-2}\right)^{{1\over N-1}},\nn
&&m_{11}=\tilde m_{22},\quad m_{22}=\tilde m_{11}.
\end{eqnarray}
For the gauge invariant monomials 
\begin{eqnarray}
 a_{12}=m_{22} , \,a_{21}=m_{11},  \quad b_{11}=\tilde m_{12}m_{21} ,\, b_{33}=m_{12}\tilde m_{21} \ , 
\end{eqnarray}
one finds the deformed geometry
\begin{eqnarray}
  b_{11}b_{33}=(a_{12}a_{21}-\mu)(a_{12}a_{21}-\tilde\mu) \ ,
\label{def}
\end{eqnarray}
where we introduced the parameters $\mu=\left(\Lambda^{3N-2}m^{N-2}\right)^{{1\over N-1}}$ 
and $\tilde\mu=\left(\tilde\Lambda^{3N-2}m^{N-2}\right)^{{1\over N-1}}$. 
%In the next section it will turn out that $\mu=\tilde\mu$.

\subsubsection{Matrix Model}
One can write a DV matrix integral description of the quiver gauge theory F-terms.
It takes the form:
\beq
Z = {1\over V}\int d\Phi_i^+ d\Phi_i^- dA_i dB_j d\tilde{A}_i d\tilde{B}_j exp
\left[-{1\over g_s} W_{tree}(\Phi_i^+, \Phi_i^- A_i,B_j, \tilde{A}_i, \tilde{B}_j)\right] \ ,
\eeq
with  $W_{tree}$ given by (\ref{W0A}), and $V$ is the volume of the groups
$SU(\hat{N_0})^2\times SU(\hat{N_0}+\hat{N_1})^2$.

For $N_0=0,N_1=N$ one can use the matrix integral 
\begin{eqnarray}
 Z=\frac{1}{V(SU(\hat{N})\times SU(\hat{N}))}\int d\Phi^{+}_{1}d\Phi^{+}_{2} e^{-\frac{m}{g_s}\,tr(\Phi^+_1\Phi^+_2)}
\ ,
\end{eqnarray}
which gives perturbatively
\beq
 {\cal F}_{\rm matrix~ model}(S) = 2\,\times\,{\cal F}_{c=1}(R_{\rm self-dual})(\mu) 
= {\cal F}_{\rm 0A}(R=1)(\mu) \ ,
\eeq
with $S=g_s \hat{N}$.
Thus, we see that
that the F-terms of the gauge theory with $N_0=0,N_1=N$ are
related to the partition function of the 0A string at the radius (\ref{topp}).
The above suggests that $\mu=\tilde{\mu}$ in (\ref{def}). 
Note, that the discussion in section 6 suggests that the same matrix model describes
the $Z_2$ orbifold of the $c=1$ bosonic string, but now  $\mu \neq\tilde{\mu}$.

\subsubsection{Quiver Gauge Theories for 0B}

One repeats the same analysis for the T-dual type 0B theory.
Again, 
the quiver gauge theory and the
 superpotential can be deduced  from the ${\hat A}_1$ quiver 
theory of the conifold. Now the $Z_2$-action takes the form
\begin{eqnarray}
\bZ_2:~~~~  \bA_2\rightarrow -\bA_2,\quad \bB_2\rightarrow -\bB_2 \ .
\end{eqnarray}
%Consider $N_0$ D3-branes and $N_1$ wrapped D5-branes.
One has

\begin{eqnarray}
  \begin{array}{ccccc}&SU(N_1)&SU(N_2)&SU(N_3)&SU(N_4)\\
    \tilde{A}_1&\bar\Box&\Box&&\\
    \tilde{B}_1&\Box&\bar\Box&&\\
    \tilde{A}_2&&\Box&\bar\Box&\\
    \tilde{B}_2&&\bar\Box&\Box&\\
    A_1&&&\bar\Box&\Box\\
    B_1&&&\Box&\bar\Box\\
    A_2&\bar\Box&&&\Box\\
    B_2&\Box&&&\bar\Box\\
    \Phi^+_1&adj.&&&\\
    \Phi^+_2&&&adj.&\\
    \Phi^-_1&&&&adj.\\
    \Phi^-_2&&adj.&&\end{array} %\ .
\end{eqnarray}
%The fields $\Phi_{1,2}^{\pm}$ arise
% from the adjoints in the $\hat A_1$ quiver and are massive, and 
As in type 0A
the $sl(2)_L\times sl(2)_R$ symmetry is broken. 

The superpotential reads
\begin{eqnarray}
  W&=&W_{0}+W_{1}\ , \nn
  W_{0}&=&\frac{m}{2}\,\left[ tr(\Phi^+_1)^2+ tr(\Phi^+_2)^2-\, tr(\Phi^-_1)^2-tr(\Phi^-_2)^2\right] \ , \nn
  W_{1}&=&-tr(A_1\Phi^+_2 B_1)-tr(A_2\Phi^+_1 B_2)-tr(\tilde A_1\Phi^+_1\tilde B_1)-tr(\tilde A_2\Phi^+_2\tilde B_2)-\nn
&&-tr(B_1\Phi^-_1 A_1)-tr(B_2\Phi^-_1 A_2)-tr(\tilde B_1\Phi^-_2 
\tilde A_1)-tr(\tilde B_2\Phi^-_2\tilde A_2)
\end{eqnarray}

Upon integrating out the massive matter one gets 
\begin{eqnarray}
%\label{suppot}
W &\sim&  tr(A_1B_1A_2B_2)-tr(\tilde A_1\tilde B_1\tilde A_2 \tilde B_2) \nn
   &-& tr(A_1\tilde B_2\tilde A_2B_1)-tr(\tilde A_1B_2 A_2\tilde B_1) \ .
\end{eqnarray}

The bifundamental fields are related to the bifundamental fields of the 0A theory  
by
\beq
  B^{\rm 0B}_i=\varepsilon_{ij} B^{\rm 0A}_j \ .
\eeq
One can see that the superpotentials match, once the massive matter is integrated out. 
This map corresponds to the T-duality of the non-critical circle line 
strings.

The analysis of the IR physics and the matrix model are the same as in the type 0A case.

\section{Discussion}

In this paper we studied the connections between four types of systems: $\hat{c}=1$
non-critical strings, 
topological B-models on quotients of the conifold, matrix models in $0+1$ and
$0+0$ dimensions and ${\cal N}=1$ supersymmetric quiver gauge theories in four dimensions.
The basic tool for identifying
these different systems was the 
commutative and associative
ring structure. 

We considered two families of non-critical strings: 
the circle line theories 0A and 0B at particular radii, and 
the super affine theories at their self-dual radii.
Based on the analysis of the ground ring and the quiver gauge theories 
we proposed these non-critical strings as equivalent
(non-perturbative) descriptions of the topological B-model on the
respective quotients of the conifold.

We discussed the superaffine and circle line models of the $0$A string
at special radii in more detail. For the superaffine theory
we derived the partition function. The analysis
of the $0$A matrix model revealed that the $\tau$-function of the
Toda hierarchy that describes the $0$A tachyon correlators can be lifted
to a two matrix integral. Following \cite{Aganagic:2003qj} 
we attempted to identify the integrable structure of the $0$A string
in the B-model on the appropriate quotient of the conifold leaving us with
many interesting open questions.
For one thing 
we constructed the deformed ring relations by indirect means.
It would be interesting to verify them.
We considered the models at special radii. It would be interesting
to analyse whether the
systems have topological descriptions at multiples of these radii
\footnote{See
section 6 for a discussion of the bosonic $c=1$ string at multiples
of the self-dual radius.}.
Another interesting direction to go would be to analyze the B-model at the
points $q \ne 0, \mu \to 0$ 
and $q \to 0$ as outlined in section \ref{Section6}.
%The case with RR charge $q\neq 0$
%seems to be more subtle, as discussed in section 6 and deserves further study.
In that context one should also strive for a better understanding
of the role of T-duality. The tachyon perturbations of the T-dual model,
or equivalently the winding mode perturbations, should match
to deformations of the fiber, which was parametrized by $u,v$ in our conventions.
Also, we have included only one RR charge. It would be interesting
to analyse the case with two RR charges different than zero.
Another generalization that deserves consideration is 
the (RR background) non-critical superstring \cite{Kutasov:1990ua}.

Note that although the non-critical strings that we studied are well defined
non-perturbatively, we established the connections between the different systems only 
perturbatively.
In the B-model there are genuine non-perturbative terms
corresponding to D-branes\footnote{For a recent study of D-brane instantons in the
topological B-model see \cite{Witten:2003nn}.}. The non-perturbative terms
in the matrix model should give non-perturbative graviphoton terms in the
effective gauge system \cite{Gopakumar:1998vy}.
%
%Such non-perturbative terms should correspond to D-brane instantons in the
%topological B-model\footnote{For a recent study of D-brane instantons in the
%topological B-model see \cite{Witten:2003nn}.},
It seems worthwhile to understand 
better the relation between the non-perturbative
contributions in the matrix model and in the B-model \cite{us}. 
It would be interesting if further understanding can be gained from the 
Lagrangian branes of the topological A-model on the quotients of the conifold, as
the S-duality between the topological A and B models suggests  
\cite{Nekrasov:2004js} and to uncover the relation
to other non perturbative
aspects of topological strings discussed recently in \cite{Okounkov:2003sp},
\cite{Iqbal:2003ds}.

\section*{Acknowledgements}

We would like to thank O. Aharony, B. Fiol, L. Horwitz,  
V. Kazakov and I. Kostov for discussions. 
This research is supported by the US-Israel Binational Science
Foundation and the TMR European Research Network.

\newpage

\appendix

\section{Leg Pole Factors for the $0$A Theory}

The leg pole factors for the deformed matrix model
have been derived in \cite{Demeterfi:1993cm} for the case $\mu=0$.
It is no more difficult to derive them for the general case,
see also \cite{Gukov:2003yp}.
As in the $c=1$ \cite{Moore:1991ir},\cite{Moore:1991sf} case we can extract the physical tachyon operators $T_p$
from the small $l$ limit of the macroscopic loop operators
\begin{equation}
\label{AppLoop1}
W\left(l,p\right) \sim \int dt e^{ipt}\mathrm{Tr} e^{-l\Phi(t)^2}=
\Gamma\left(-p\right)l^p T_p \ ,
\end{equation}
see also \cite{Moore:1991zv}.
If we consider a correlator of the loop operators
\begin{equation}
\left<\prod_i^nW\left(l_i,p_i\right)\right>_c=\int \left(
\prod_id\lambda_ie^{-l_i\lambda_i} \right)\left<\prod_i\rho\left(\lambda_i,p_i\right)\right>_c \ ,
\end{equation}
introduce the classical time $\tau$ through
\begin{equation}
\lambda^2=\mu+\sqrt{q^2+4\mu^2-\frac{1}{4}}\cosh 2 \tau \ ,
\end{equation}
and consider the limit $\tau \to \infty, \; l_i \to 0$ in (\ref{AppLoop1})
we find that
\begin{equation}
\left<\prod_i^nW\left(l_i,p_i\right)\right>_c \sim 
\prod_i \left(\Gamma\left(-p_i\right) l^{p_i} 2^{-p_i/2}
\left(q^2+4\mu^2-\frac{1}{4}\right)^{p_i/4}\right)\mathcal{R}\left(p_k\right) \ ,
\end{equation}
where the part $\mathcal{R}(p_k)$ is given by products of the
reflection coefficients according to the rules of \cite{Moore:1991zv}.
If we denote by $V_p$ the tachyon vertex operators normalized according to
the conventions of this paper and by $T_p$ the physical tachyon vertex operators we
have the to use the following translation rules
\begin{equation}
\left<\prod_{i=1}^nT_{p_i}\right>=\mu^{2-n}\prod_{i=1}^n\left(
\frac{q^2+4\mu^2-1/4}{4}\right)^{\left|p_i\right|/4}
\left<\prod_{i=1}^nV_{p_i}\right> \ .
\end{equation}

\end{document}